\documentclass{revtex4-2}
\usepackage[latin9]{inputenc}
\usepackage{amsmath}
\usepackage{amssymb}
\usepackage{graphicx}
\usepackage[unicode=true,
 bookmarks=false,
 breaklinks=false,pdfborder={0 0 1},colorlinks=false]
 {hyperref}
\hypersetup{
 colorlinks,linkcolor=red,citecolor=green}

\makeatletter

\providecommand{\tabularnewline}{\\}

\usepackage[toc]{appendix}
\usepackage{epsfig}
\graphicspath{./}

\newcommand{\be}{\begin{equation}}
\newcommand{\ee}{\end{equation}}
\newcommand{\re}[1]{(\ref{#1})}

\usepackage[normalem]{ulem}
\usepackage[dvipsnames]{xcolor}
\renewcommand{\sout}{\bgroup \color{red} \ULdepth=-.5ex \ULset}

\usepackage{tikz-feynman}
\usetikzlibrary{shapes,arrows,positioning,automata,backgrounds,calc,er,patterns}
\tikzfeynmanset{compat=1.0.0, every vertex={dot,minimum size=2pt}}

\makeatother

\begin{document}
\title{Gluons, light and heavy quarks and their interactions in the instanton
vacuum}
\author{M.\,Musakhanov
}
\affiliation{National University of Uzbekistan, Tashkent 100174, Uzbekistan}
\begin{abstract}
The interactions of colorful particles with QCD vacuum instantons
are controlled by the instanton size $\rho$ and inter-instanton distance
$R$, which are the main parameters of Instanton Liquid Model (ILM)
of the QCD vacuum. Lattice, theoretical and phenomenological estimates
show that the values of these parameters are $\rho\approx1/3$~fm
and $R\approx1$~fm, and the corresponding packing parameter
$\kappa=\rho^{4}/R^{4}\approx0.01$. The strength of the light quark-instanton
interaction is sizable and close to that of the gluon-instanton one.
They are related to the dynamical light quark mass $M_{q}$ and dynamical
gluon mass $M_{g}$, which are given by  $M_{q}\approx M_{g}\approx360$~MeV$\sim\kappa^{1/2}\rho^{-1}$.
On the other hand, the strength of the heavy quark-instanton interaction
is weak, and is determined by the direct instanton contribution to
the heavy quark mass $\Delta M_{Q}^{{\rm dir}}\approx70$~MeV$\sim\kappa\rho^{-1}.$
So, the instantons are responsible for mutual interactions among colored
particles which are crossing the field of the same instanton, like
for example t'Hooft-like interactions of $N_{f}$ light quarks ($N_{f}$
is the light quark flavors number). The light quark propagators in
the instanton field have zero modes, which give dominant contributions.
Within ILM we are able to derive the light quarks determinants and
the light quarks partition function. These tools  perfectly describe
the light quark physics and its most important and basic phenomena
- the spontaneous breaking of the chiral symmetry (SBChS). This allows
to understand in details the effective low-energy Chiral Perturbation
Theory (ChPT) in terms of quark degrees of freedom, and make unambiguous
predictions of the Gasser-Leuteyller low-energy coupling constants.
We conclude that ILM is a good framework for the light hadrons physics
calculations. This framework allows us to find the properties of light
and heavy quarks interactions  by accounting of the light quarks determinant
in ILM heavy quark correlators. ILM provides the framework for the
calculation of SBChS traces in light-heavy quarks systems. As an example,
we are considering the process of pions emission by excited heavy
quarkonium states. We found a sizable corrections ( $\sim20\%$) to
the widely used dipole approximation for the process $\psi(2S)\rightarrow J/\psi\,\pi^{+}\pi^{-}$.
So, we need to find heavy $Q\bar{Q}$ quarkonia spectra and their
wave functions. Here we have interplay of two scales: short ($\sim(M_{Q}v)^{-1}\leq0.15$~fm)
perturbative QCD and large ($\sim(M_{Q}v^{2})^{-1}\sim0.5$~fm) nonperturbative
QCD scales, where  $v$ is the velocity in $Q\bar{Q}$. We calculate
the heavy quarks correlators with perturbative corrections within
ILM. We find  almost complete mutual cancellation of the ILM direct
$O(\kappa)$ instanton and ILM perturbative $O(\kappa^{1/2}\alpha_{s})$
contributions to the heavy quark correlators.
\end{abstract}
\maketitle

\section{Introduction}

\label{intro}

The QCD vacuum is quite nontrivial non-perturbative quantum state
characterized by the nonvanishing gluon and light quark condensates,
which are well described within the Instanton Liquid Model (ILM) (see
reviews~\cite{Schafer:1996wv,shuryak2018,Diakonov:2002fq}). The
most essential assumption of ILM is the applicability of the semiclassical
approximation, which means that the effective strong coupling $\alpha_{s}$
is rather small. The ILM is characterized by only two parameters:
the average instanton size $\rho\approx0.3\,{\rm fm}$ and the average
inter-instanton distance $R\approx1\,{\rm fm}$. These essential parameters
were suggested by Shuryak~\cite{Shuryak:1981ff} (see also~\cite{Schafer:1996wv,shuryak2018})
and were estimated phenomenologically in~\cite{Diakonov:1983hh}
(see also~\cite{Diakonov:2002fq}). These values are in agreement
with lattice measurements \cite{Chu:1994vi,Negele:1998ev,DeGrand:2001tm,Faccioli:2003qz},
and lead to a comfortable value of the packing parameter $\kappa=\rho^{4}/R^{4}\sim0.01$,
which characterizes the fraction of the 4D space occupied by instantons.
Such small value of $\kappa$ simplifies very much the calculations
within ILM.

One of the most prominent achievements of the Instanton Liquid Model
 is the correct description of the spontaneous breaking of the chiral
symmetry (SBChS), which is responsible for properties of light quark
hadrons and nuclei (see reviews~\cite{Schafer:1996wv,shuryak2018,Diakonov:2002fq}).
The SBChS is due to specific nonperturbative properties of the QCD
vacuum,  and is a very important object of investigations by methods
of Nonperturbative Quantum Chromo Dynamics (NPQCD). In the instanton
picture the SBChS occurs due to the delocalization of single-instanton
quark zero modes in the instanton medium, which is possible since
current light quark mass $m_{q}\ll1/\rho$. All properties of light
quark physics in ILM are controlled by a rather large value of light
quark-instanton interaction effective strength, which is proportional
to  dynamical light quark mass $M_{q}\sim\kappa^{1/2}\rho^{-1}$.
At typical values of the ILM parameters given above, this mass is
given by $M_{q}\approx360$~MeV. In contrast to this, the heavy quark
physics is controlled mainly by the heavy quark mass $M_{Q}\gg\rho^{-1}$,
with rather small influence of the instantons, since heavy quark-instanton
interaction strength is given by $\Delta M_{Q}^{{\rm dir}}\approx70$~MeV$\sim\kappa\rho^{-1}$.


In this context, we expect that essential properties of open heavy
flavor hadrons seem to be strongly governed by such light quark properties
as the phenomenon of spontaneous breaking of the chiral symmetry,
which should be taken into account in an appropriate way. Consequently,
the traces of the nonperturbative dynamics may be essential in the
open heavy flavor systems not only for their decay modes into the
lighter hadrons, but also for studies of their static properties.
Similar effects are expected in the decay processes of exited heavy
quarkonia, with emission of light hadrons.

In contrast, the heavy quarkonia spectrum and their wave functions could be affected by both perturbative and non-perturbative interaction forces~\cite{Bali:2000gf,Brambilla:2001fw}. In ILM we have to take into account the $Q\bar Q$ perturbative potential  with the modification of gluon properties due to presence of instantons in QCD vacuum together with direct instanton generated one ~\cite{Musakhanov:2020hvk}.
We have to note that, instanton media induced dynamical gluon mass
$M_g\approx  360$~MeV$\sim \kappa^{1/2}\rho^{-1}$, which tells us that
perturbative gluon-instanton interaction strength is the same as for light quarks and is
large.

The overview of current situation is also one of the aims of the present
review and will be done on a basis several recent studies\,\cite{Diakonov:1989un,Turimov:2016adx,Yakhshiev:2018juj,Musakhanov:2020hvk}
developed in the framework of instanton liduid model (ILM) of QCD
vacuum. We organize our review as follows: in the subsection~\ref{ILM}
of the Introduction we briefly remind the main concepts of the instanton
liquid model (ILM) for QCD vacuum and its phenomenological parameters.
The section~\ref{lightquarks} is devoted to discussion of the physics
of the light quarks in ILM.  In the subsection~\ref{mesons} of that
section we focus on the derivation of chiral light mesons effective
action in the presence of external fields for the case of light quarks
number $N_{f}=2$. The most essential part of our review is the section~\ref{Qq},
in which we consider the heavy and light quarks systems in ILM. In
its subsection~\ref{QI} we calculate the strength of heavy-quark-instanton
interaction, and in the following subsection~\ref{Qqinteractions}
we consider the instanton-generated heavy and light quark interactions
in ILM. The subsection~\ref{QQq} is devoted to the heavy quarkonium
and light quarks interactions in ILM, where it is derived the effective
action corresponding to the processes of pions emission by heavy quarkonium.
In the subsection~\ref{QQsizes} we compare the heavy quarkonium
states and instanton sizes to clarify in the following subsection~\ref{QQpipi-standard}
the mechanism of the $(Q\bar{Q})_{n'}\rightarrow(Q\bar{Q})_{n}\,\,\pi\pi$
processes. From our estimates we have a sizable correction $\sim20\%$
to the standard dipole approximation for the case of $\psi'\rightarrow J/\psi\pi\pi$
process. In order to make precise calculations of this and similar
processes, we have to calculate the quarkonium spectrum and wave functions
within ILM. It requires the calculations of heavy quark correlators
with perturbative $O(\alpha_{s})$ corrections in ILM, which are considered
in the section~\ref{QQCor}. In the subsection~\ref{gluons} the
gluon propagator is calculated within ILM, which is used in the subsection
~\ref{Qprop} for the calculations of ILM perturbative corrections
to the heavy quark mass and in the subsection~\ref{QQpotential}
is applied to the calculations of heavy quarks $Q\bar{Q}$ singlet
potential with perturbative corrections in ILM.

In the section~\ref{summary} we give an overview of the ILM results
and formulate the direction of the following research.

\subsection{Instanton liquid model (ILM) for QCD vacuum and its parameters}

\label{ILM}

The instanton is one of the possible manifestations of the rich topological
structures of QCD vacuum. Mathematically, an instanton is a topologically
non-trivial classical solution of the Yang-Mills equations in the
4-dimensional Euclidean space, and mathematically corresponds to the
mapping of $SU(2)$ part of internal $SU(N_{c})$ color space to the
appropriate $SU(2)$ part of the 4-dimensional Euclidean space. 

\begin{figure}[hbt]
\includegraphics[scale=0.7]{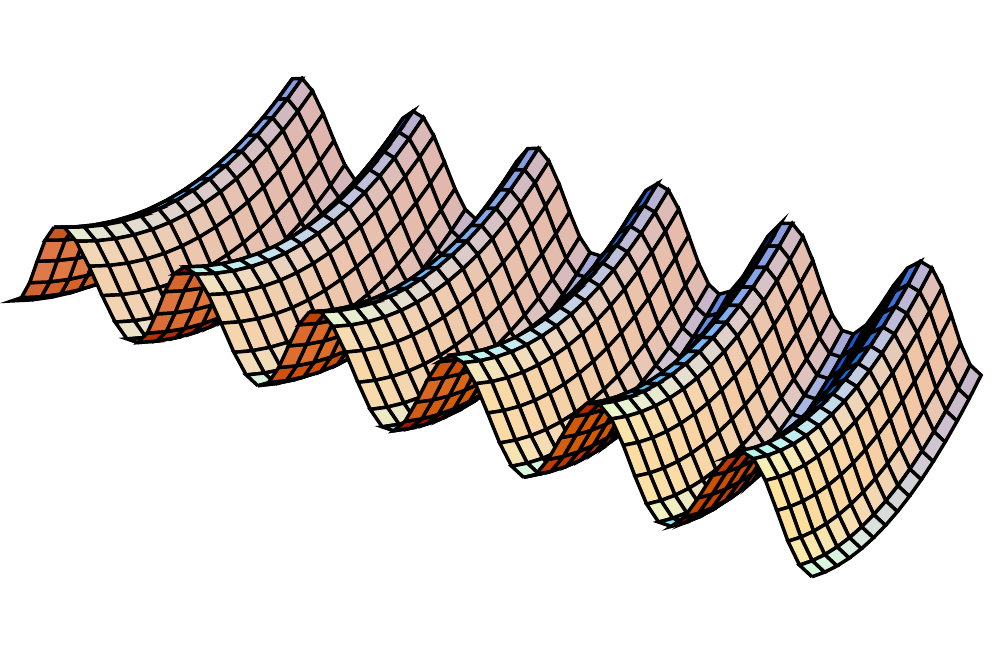},
\includegraphics[scale=0.7]{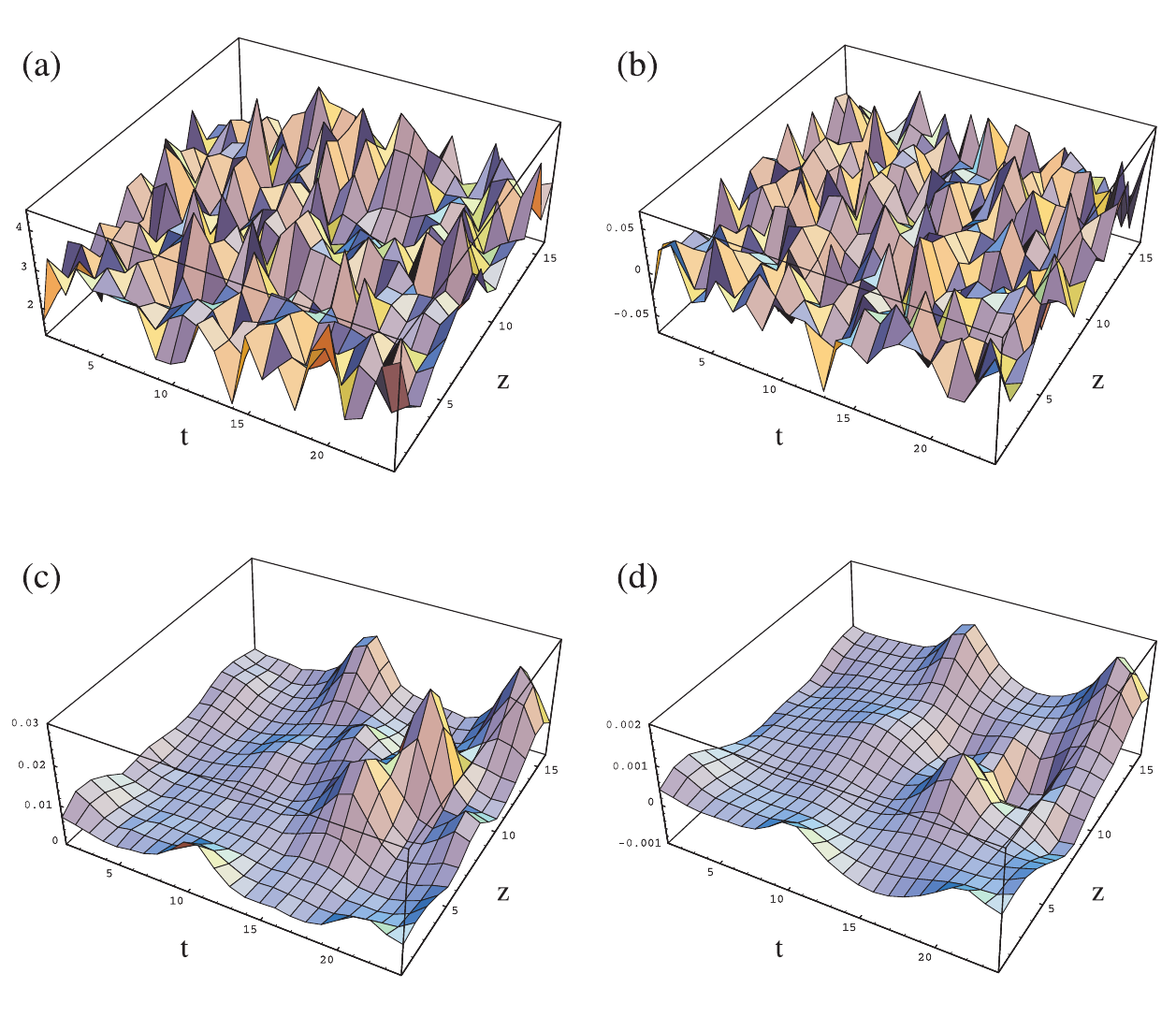}
\caption{\textbf{Left plot:} Schematic illustration of the periodic dependence
of potential energy on Chern-Simons number $N_{{\rm CS}}$, and mild
(oscillator-like) growth in all other directions in functional space~\cite{FJR1976,Jackiw:1976pf}.
\textbf{Right plot:} \textit{Upper row}: a typical full configuration
of the gluon field from lattice simulations~ in the $(z,t)$ hyperplane
with $(x,y)$ fixed. \textit{Lower row}: the same after smearing of
zero-point oscillations clearly shows 3 instantons and 2 anti-instantons.
\textit{Left column}: action density. \textit{Right column}: topological
charge density~\cite{Negele:1998ev}. }
\label{periodic} 
\end{figure}

The potential part of the classical Yang-Mills action has  periodic
dependence on the collective coordinate, which is called the Chern-Simons
(CS) number $N_{{\rm CS}}$. The classical vacuum, being defined as
minimum energy state, is infinitely degenerated  set of solutions
which differ by this choice of $N_{{\rm CS}}=n\in\mathbb{Z}$, as
could be seen from the Fig.~\ref{periodic}. It can be also considered
as the lowest energy quantum state of the one-dimensional periodic
crystal along the CS coordinate~\cite{FJR1976,Jackiw:1976pf}. In
this context, the instanton might be considered as a tunneling path,
which joins adjacent classical vacua with Chern-Simons indices $n$
and $n+1$, while the anti-instanton corresponds to transition in
opposite direction~\cite{Belavin:1975fg}. 

\begin{figure}[hbt]
\begin{centering}
\includegraphics[scale=0.75]{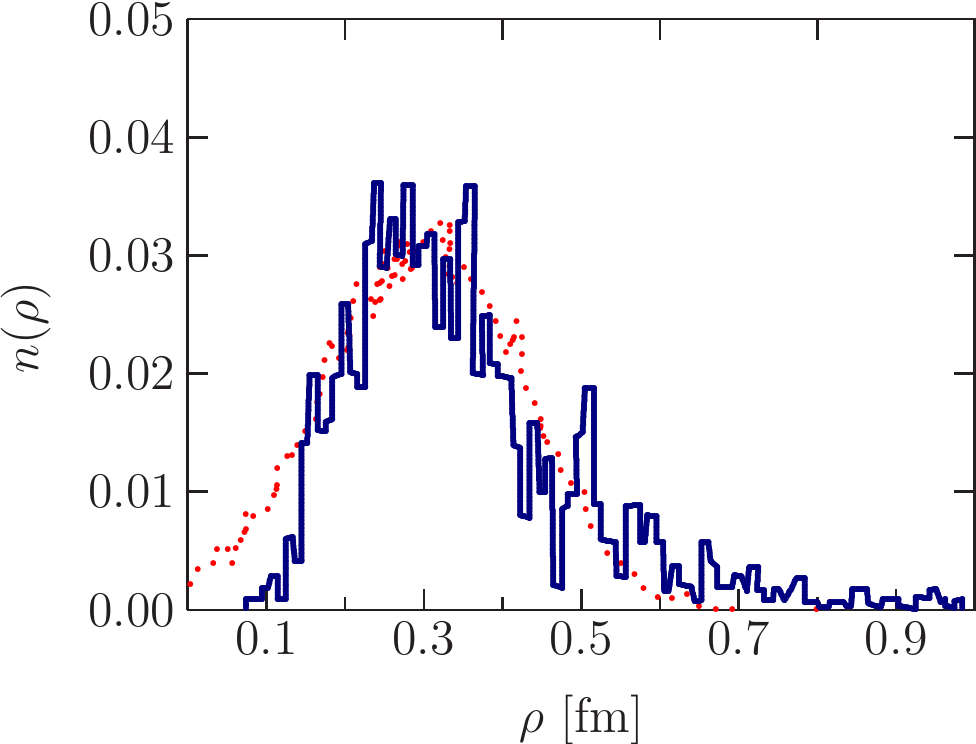}
\par\end{centering}
\caption{The instanton size distribution function $n(\rho)$. The dots correspond
to the calculations in the framework of ILM, while the continuous
lines correspond to the lattice simulations~\cite{Millo:2011zn}.}
\label{instantonsize} 
\end{figure}

In what follows we will use a shorthand notation $\xi_{I},\xi_{\bar{I}}$
for the collective coordinates of the instantons and antiinstantons:
the position in 4-dimensional Euclidean space $z_{I}$, the instanton
size $\rho_{I}$ and the unitary SU$(N_{c})$ color orientation  matrix
$U_{I}$, which itself depends on $4N_{c}$ variables, where $N_{c}$
is the number of colors.  Hereafter, sometimes we will drop the subscripts
$I,\bar{I}$ for the sake of brevity, when this does not cause additional
confusion.

There are two main parameters in ILM -- the average instanton size
$\bar{\rho}$ and the inter-instanton distance $R$. The latter describes
the density of instanton media $N/V\equiv1/R^{4}$, where $N$ is
the total number of instantons.  Phenomenologically this density is
related to the gluon condensate as~\cite{Shifman:1978bx} 
\begin{equation}
\frac{N}{V}\simeq\frac{1}{32\pi^{2}}\langle F_{\mu\nu}^{a}F_{\mu\nu}^{a}\rangle\simeq(200\,{\rm MeV})^{4}.
\end{equation}
which allows to obtain an estimate $R\simeq1\,{\rm fm}$.The instanton
size $\rho$ effectively sets the scale, which controls the value
of the renormalized strong coupling $\alpha_{s}$~\cite{tHooft:1976snw,Callan:1977gz}.
In one-loop approximation ( $\overline{{\rm MS}}$ scheme) we have
\begin{eqnarray}
\frac{2\pi}{\alpha_{s}(\rho)}=b_{1}\ln\frac{1}{\Lambda_{\overline{MS}}\,\rho},\,\,\,b_{1}=\frac{11}{3}N_{c}-\frac{2}{3}N_{f}\label{alphas}
\end{eqnarray}
where $N_{c}$ is the introduced earlier  number of colors, and $N_{f}$
is the number of light flavors. At $1/\rho=600$~MeV, using $\Lambda_{\overline{MS}}=100$~MeV,
we may obtain a rather large value$\alpha_{s}(\rho)=0.39.$

Due to conformal symmetry of classical QCD, a single instanton may
have any size. However, due to quantum corrections, as we approach
$\rho\Lambda_{\overline{MS}}\sim1$, the semiclassical approximation
becomes absolutely inapplicable. In the  instanton-antiinstanton medium,
due to mutual interactions, both parameters $R$ and $\rho$ are close
to their average values $\bar{R}\approx1$~fm, $\bar{\rho}\approx0.33$~fm
values, which  were found by the application of Feynman variational
method to this problem by Diakonov and Petrov~\cite{Diakonov:1983hh}
(see also reviews~\cite{Diakonov:2002fq,Schafer:1996wv,shuryak2018}),
and further confirmed by the lattice simulations of the QCD
vacuum~\cite{Chu:1994vi,Negele:1998ev,DeGrand:2001tm,Faccioli:2003qz}
 (see Fig.~\ref{periodic}).
A detailed studies of the instanton size distribution $n(\rho)$ was
 made in~\cite{Millo:2011zn}, and is shown in Fig.~\ref{instantonsize},
where the results in framework of ILM model are also given for  comparison.
We can see that at the relatively large values of instanton sizes
$\rho\gtrsim R$, where more intensive overlapping of instantons might
happen, the distribution function $n(\rho)$ is strongly suppressed.
A rather narrow peak of the distribution is localized around $\rho\simeq0.35$\,fm,
which corresponds to the average size $\bar{\rho}$. Therefore, for
practical calculations we can disregard distribution of instantons
by size, and effectively replace them all with average-size instantons.
In what follows we will use this scheme and disregard possible corrections
due to finite width of the distribution $n(\rho)$, assuming $\rho=\bar{\rho}$.
The diluteness of the instanton gas and smallness of the packing fraction
$\kappa=\bar{\rho}^{4}/R^{4}\sim0.01$ justifies the use of simple
sum-ansatz for the total instanton field $A(\xi)=\sum_{i}A_{i}(\xi_{i})$,
which is expressed in terms of the single instanton solutions $A_{i}(\xi_{i})$
Using the typical values of the ILM parameters $R=1$~fm, $\bar{\rho}=1/3$~fm,
we can estimate the QCD vacuum energy density, which takes the nonzero
value, $\epsilon\approx-500\,{\rm MeV/fm^{3}}$\,\cite{Schafer:1996wv,shuryak2018}.

Finally, we would like to mention  that the large-size tail of the
instanton distribution $n(\rho)$ becomes important in the confinement
regime of QCD. Here in order to take instanton phenomena more accurately,
we should replace Belavin-Polyakov-Schwarz-Tyupkin instantons by Kraan-vanBaal-Lee-Lu
instantons~\cite{Kraan:1998kp,Kraan:1998pm,Lee:1998bb} described
in terms of dyons. In such a way, we get a natural extension of the
instanton liquid model, the so-called liquid dyon model (LDM)\,\cite{Diakonov:2009jq,Liu:2015ufa,Liu:2015jsa}.
This model allow to reproduce the confinement{--}deconfinement
phases. The small size instantons in this approach can still be described
in terms of their collective coordinates. For comparison, the average
size of instantons in liduid dyon model is $\bar{\rho}_{{\rm LDM}}\approx0.5\,{\rm fm}$
\cite{Diakonov:2009jq,Liu:2015ufa,Liu:2015jsa}, slightly larger than
the above-mentioned  estimate in the instanton liquid model. 

\section{Light quarks in ILM}

\label{lightquarks}

As discussed earlier in Introduction, one of the most prominent achievements
of the instanton vacuum model is the correct description of the spontaneous
breaking of the chiral symmetry (SBChS), which is responsible for
properties of most hadrons and nuclei ~\cite{Leutwyler:2001hn} and
 is due to specific nonperturbative properties of QCD vacuum. In the
instanton picture SBChS occurs due to the delocalization of the single-instanton
quark zero modes in the instanton medium. 
\begin{figure}[hbt]
\includegraphics[scale=0.6]{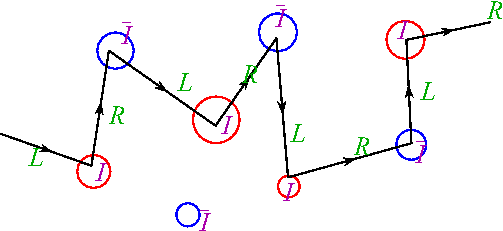}\,\,\,\,\,\,\,\includegraphics[scale=0.4]{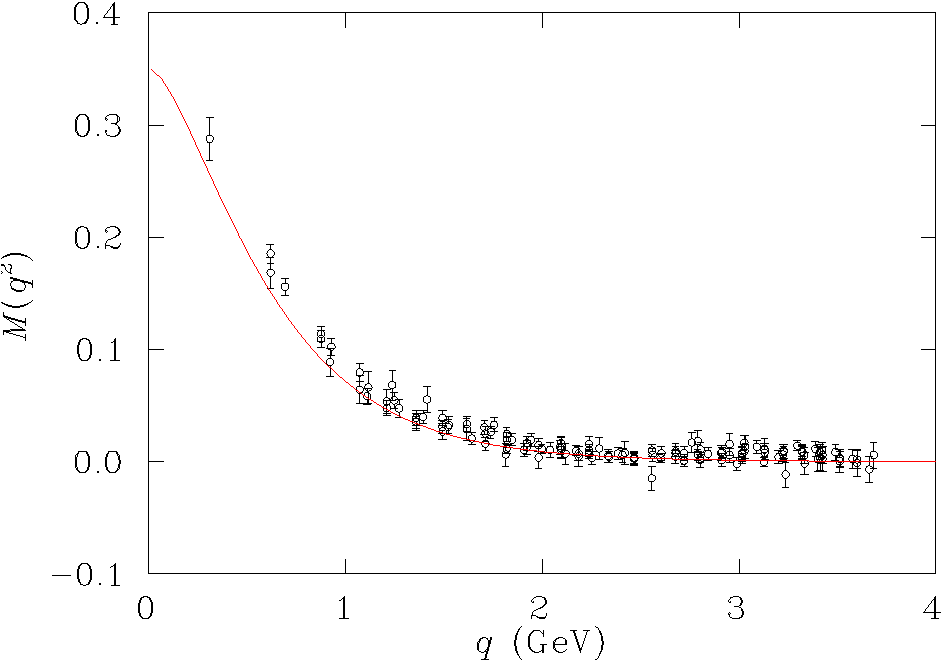}
\caption{\label{fig:3}\textbf{Left plot:} Schematic illusration of quark rescatterings
in the field of individual instantons and antinistantons which leads
to SBChS. \textbf{Right plot:} Comparison of the dynamical quark mass
$M_{q}(q)$ from ILM and lattice calculations~\cite{Bowman:2005vx}.}
\end{figure}

The integration over the light quark degrees of freedom in the partition
function leads to the  determinant of the light quark propagator $S$
evaluated in the background field of instanton liquid. We may split this
determinant into the low- and high-frequency parts according ${\rm Det}={\rm Det}_{\mathrm{low}}{\rm Det}_{\mathrm{high}}$.
The high-energy part ${\rm Det}_{\mathrm{high}}$ is responsible mainly
for the perturbative coupling renormalization, for this reason in
what follows we disregard it  and concentrate on the evaluation of
the nonperturbative contribution ${\rm Det}_{\mathrm{low}}$, which
is responsible for the low-energy domain. As was suggested in~~\cite{tHooft:1976snw,Lee:1979sm,Diakonov:1985eg,Diakonov:1995qy},
we may start evaluation with the zero-mode approximation for the light
quark propagator of a quark in the field of the isolated $i$-th instanton,
\begin{equation}
S_{i}=\frac{1}{\rlap{/}{p}+\rlap{/}{A}_{i}+im}=\frac{1}{\rlap{/}{p}}+\frac{|\Phi_{i,0}\rangle\langle\Phi_{i,0}|}{im}.\label{SiDP}
\end{equation}
While this approximation is good for small values of the current quark
mass $m$, we need to extend it beyond the chiral limit as proposed
in our previous works~ \cite{Musakhanov:1998wp,Musakhanov:2002vu,Musakhanov:2002xa,Kim:2004hd,Kim:2005jc,Goeke:2007nc}
as follows: 
\begin{equation}
S_{i}=S_{0}+S_{0}\rlap{/}{p}\frac{|\Phi_{0i}\rangle\langle\Phi_{0i}|}{c_{i}}\rlap{/}{p}S_{0},\label{Si}
\end{equation}
where 
\begin{equation}
c_{i}=-\langle\Phi_{0i}|\rlap{/}{p}S_{0}\rlap{/}{p}|\Phi_{0i}\rangle=im\langle\Phi_{0i}|S_{0}\rlap{/}{p}|\Phi_{0i}\rangle=im\langle\Phi_{0i}|\rlap{/}{p}S_{0}|\Phi_{0i}\rangle.
\end{equation}
The approximation given in Eq.(\ref{Si}) allows us to get correct
projection of $S_{i}$ on the would-be zero-modes in presence of finite
quark mass $m$, 
\begin{equation}
S_{i}|\Phi_{0i}\rangle=\frac{1}{im}|\Phi_{0i}\rangle,\,\,\,\langle\Phi_{0i}|S_{i}=\langle\Phi_{0i}|\frac{1}{im}.
\end{equation}

As we mentioned earlier, in dilute liquid approximation the instanton
ensemble might be represented as a superposition of gluonic fields
of individual instantons, for this reason we can represent the total
quark propagator ${S}$ in the presence of the whole instanton ensemble
$A$ in terms of the quark propagators of individual instantons. Physically,
this corresponds to resummation of a multi-scattering series, shown
schematically in the left panel of the Figure~\ref{fig:3}. 
 Making a few technical steps~\cite{Goeke:2007nc}, we may represent
this determinant in terms of the low-energy contributions of individual
instantons. In what follows we will focus on evaluation of the partition
function, which depends on the external quark sources $\eta,\bar{\eta}$,
for this reason from now on we will incorporate them in our consideration.
 The fermionized representation of the low-frequencies light quark
determinant in the presence of these quark sources $\bar{\eta},\,\eta$
is given by  
\begin{eqnarray}
 &  & {\rm Det}_{{\rm low}}\exp\left(-\eta^{+}S\eta\right)=\nonumber \\
 &  & =\int\prod_{f}D\psi_{f}D\psi_{f}^{\dagger}\exp\int\sum_{f}\left(\psi_{f}^{\dagger}(\hat{p}\,+\,im_{f})\psi_{f}+\psi_{f}^{\dagger}\eta_{f}+\eta_{f}^{+}\psi_{f}\right)\prod_{f}\prod_{\pm}^{N_{\pm}}V_{\pm,f}[\psi^{\dagger},\psi,\xi_{\pm}],\label{part-func}
\end{eqnarray}
where the vertices
\begin{eqnarray}
V_{\pm,f}[\psi^{\dagger},\psi,\xi_{\pm}]=i\int d^{4}x\left(\psi_{f}^{\dagger}(x)\,\hat{p}\Phi_{\pm,0}(x;\xi_{\pm})\right)\int d^{4}y\left(\Phi_{\pm,0}^{\dagger}(y;\xi_{\pm})(\hat{p}\,\psi_{f}(y)\right)\label{V}
\end{eqnarray}
correspond to interaction of the light quarks with individual instantons.

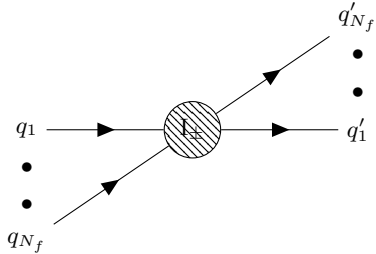
\begin{figure}[!ht]
\centering{}\tikz{\begin{feynman} \vertex (i1){}; \vertex(q1)
[below=1cm of i1]{$q_{1}$}; \vertex(m1) [below=0.5cm of
q1]{$\bullet$}; \vertex(n1) [below=0.5cm of m1]{$\bullet$};
\vertex(o1) [below=0.5cm of n1]{$q_{N_{f}}$}; \vertex(c)[blob]
[right=2.2cm of q1]{${\rm I_{\pm}}$}; \vertex(q2) [right=2.2cm
of c]{$q'_{1}$}; \vertex(n2) [above=0.5cm of q2]{$\bullet$};
\vertex(m2) [above=0.5cm of n2]{$\bullet$}; \vertex(o2) [above=0.5cm
of m2]{$q'_{N_{f}}$}; \diagram*{ (q1) --[fermion] (c)
--[fermion] (q2), (o1) --[fermion] (c) --[fermion](o2);
};\end{feynman}} \caption{Instanton-generated light quarks  interaction $W_{\pm}[\psi^{\dagger},\psi]$
with $2N_{f}$-legs.}
\label{figq} 
\end{figure}

The averaging over the collective coordinates $\xi_{i,\pm}$ in~(\ref{part-func},~\ref{V})
is straightforward, since the low density of the instanton liquid
allows to average over positions and orientations of the individual
instantons independently from each other. This process leads to the
light quark partition function $Z[\eta,\eta^{\dagger}]$.
We see that
\begin{equation}
Y_{\pm}[\psi^{\dagger} ,\psi]= \int d\xi_\pm \prod_{f}V_{\pm,f}[\psi^{\dagger},\psi,\xi_{\pm}]\label{eq:Y}
\end{equation}
is instanton-generated light quarks $2N_f$-legs interaction represented in Fig.~\ref{figq}. 

The nonlocal form-factor in the interaction vertex~(\ref{V},~\ref{eq:Y})
is completely defined by the quark zero-mode.  For further evaluations
it is convenient to raise the vertices $Y_{\pm}$ into the argument
of exponent, using the so-called exponentiation with the help of Stirling-like
formula 
\begin{eqnarray}
Y_{\pm}^{N_{\pm}}=\int d\lambda_{\pm}\exp(N_{\pm}\ln\frac{N_{\pm}}{\lambda_{\pm}}-N_{\pm}+\lambda_{\pm}Y_{\pm}),\label{lambda}
\end{eqnarray}
where $\lambda_{\pm}$ play a role of the dynamical coupling constant,
defined by saddle-point condition at the integral \re{lambda}. The
validity of the formula~(\ref{lambda}) is controlled by the large
number of instantons $N_{\pm}\gg1$. The partition function of the
light quarks~(\ref{part-func}) at $N_{f}=1$ and $N_{\pm}=N/2$
is exactly given by 
\begin{eqnarray}
 &  & Z\left[\eta,\eta^{\dagger}\right]=e^{-\eta^{+}\left(\hat{p}\,+\,i(m+M(p))\right)^{-1}\eta}\exp\left[{\rm Tr}\ln\left(\hat{p}\,+\,i(m+M_{q}(p))\right)+N\ln\frac{N}{2\lambda}-N\right],\label{Z}\\
 &  & N={\rm Tr}\frac{iM_{q}(p)}{\hat{p}\,+\,i(m+M_{q}(p))},\,\,\,M_{q}(p)=M_{q}F^{2}(p),\,\,\,\,M_{q}=\frac{\lambda}{N_{c}}(2\pi\rho)^{2},\label{M}
\end{eqnarray}
where the form-factor $F(p)$ is related to the Fourier-transform
of the zero-mode. The  dynamical quark mass $M_{q}(p)$  defined in
the Eq.~(\ref{M}) is the instanton-induced nonperturbative contribution
to the dynamical quark mass, whose dependence on the quark momentum
is shown in the right panel of the Figure~\ref{fig:3}. From this
equation we can see that it simultaneously determines the strength
of the light quark-instanton interaction, which is parametrically
given by $M_{q}\sim\kappa^{1/2}\rho^{-1}$. Its numerical value might
be found solving the first equation in~(\ref{M}), and at typical
values of $\bar{\rho}\approx1/3$~fm, $R\approx1$~fm we may get
$M_{q}\approx360$~MeV. For number of flavors $N_{f}>1$, using
the saddle-point approximation, in the leading order we may get the
same value of $M_{q}$.

\subsection{Light mesons effective action at $N_{f}=2$ with nonleading order
(NLO) contributions \protect \\
 and Chiral Perturbation Theory (ChPT) couplings }
\label{mesons} 
The partition function in the presence of external
fields provides a straightforward way for the calculations of various
correlators. This function for the two light flavors has been evaluated
in the framework of ILM  in our previous paper~\cite{Goeke:2007bj}.
We found that the light quarks partition function $Z[v,a,s,p,m]$
in the presence of external vector $v_{\mu}$, pseudovector $a_{\mu}$,
scalar $s$ and pseudoscalar $p$ fields is given by: 
\begin{eqnarray}
 &  & \tilde{Z}[v,a,s,p,m]=\int\prod_{f}D\psi_{f}D\psi_{f}^{\dagger}\exp\left(\int d^{4}x\sum_{f,g}\psi_{f}^{\dagger}(\hat{p}\,+\,\hat{V}\,+\,im)_{fg}\psi_{g}\right)\prod_{\pm}^{N_{\pm}}\tilde{Y}_{\pm}[\psi^{\dagger},\psi],\nonumber \\
 &  & \tilde{Y}_{\pm}[\psi^{\dagger},\psi]=\int d\xi_{\pm}\prod_{f}\tilde{V}_{\pm,f}[\psi^{\dagger},\psi],\label{tildeY}
\end{eqnarray}
where 
\begin{eqnarray}
\tilde{V}_{\pm}[\psi^{\dagger},\psi]=\int d^{4}x\left(\psi^{\dagger}(x)\,\bar{L}^{-1}(x,z_{\pm})\,\hat{p}\Phi_{\pm,0}(x;\xi_{\pm})\right)\int d^{4}y\left(\Phi_{\pm,0}^{\dagger}(y;\xi_{\pm})(\hat{p}\,L^{-1}(y,z_{\pm})\psi(y)\right),\label{tildeV}
\end{eqnarray}
$\tilde{Y}_{\pm}[\psi^{\dagger},\psi]$ is the t'Hooft-like non-local
interaction term with $N_{f}$ pairs of quark legs in the presence
of the external fields, and  $L_{i}$ is the gauge connection defined
as a path-ordered exponent 
\begin{eqnarray}
L_{i}(x,z_{i})={\rm P}\exp\left(i\int_{z_{i}}^{x}dy_{\mu}(v_{\mu}(y)+a_{\mu}(y)\gamma_{5})\right),\,\,\,\bar{L}_{i}(x,z_{i})=\gamma_{4}L_{i}^{\dagger}(x,z_{i})\gamma_{4};\label{transporter}
\end{eqnarray}
the variable $z_{i}$ denotes the instanton or antiinstanton position.
After integration over collective coordinates and exponentiation,
the partition function may be reduced to the form : 
\begin{eqnarray}
 &  & \tilde{Z}=\int d\lambda_{+}d\lambda_{-}D\psi^{\dagger}D\psi\exp\left[-\left(N_{\pm}\ln\frac{K}{\lambda_{\pm}}-N_{\pm}+\psi^{\dagger}(i\hat{\partial}+\hat{V}+im)\psi+\lambda_{\pm}\tilde{Y}_{2}^{\pm}\right)\right];\nonumber \\
 &  & \tilde{Y}_{2}^{\pm}=\frac{2N_{c}-1}{2N_{c}(N_{c}^{2}-1)}\left(\det_{f}J^{\pm}+\frac{1}{8N_{c}-4}\det_{f}J_{\mu\nu}^{\pm}\right),\nonumber \\
 &  & J^{\pm}=\psi^{\dagger}\bar{L}\frac{1\pm\gamma_{5}}{2}L^{-1}\psi,\,\,J_{\mu\nu}^{\pm}=\psi^{\dagger}\bar{L}\sigma_{\mu\nu}\frac{1\pm\gamma_{5}}{2}L^{-1}\psi,\label{tildeZ}
\end{eqnarray}
where the determinant is taken over the (implicit) flavor indices,
and $K$ is some inessential constant introduced to make the argument
of logarithm dimensionless. From (\ref{tildeZ}) we can clearly see
that the contribution of the tensor terms is just a $1/N_{c}$-correction.
For the sake of simplicity we will postpone consideration of the tensor
terms contribution and concentrate on the first term in~(\ref{tildeZ}).
The expression $\det_{f}J^{\pm}$ includes a product of $2N_{f}$
quark operators, which might be represented in a more convenient form
using the bosonization of the interaction term $H_{\pm}[\psi^{\dagger},\psi]$.
For the sake of simplicity we will consider  only the $N_{f}=2$ case,
for which bosonization is an exact procedure, and take $N_{+}=N_{-}$,
as discussed before. We get for the partition function 
\begin{eqnarray}\nonumber
 &  & \tilde{Z}[v,a,s,p,m]=\int d\lambda D\Phi D\psi^{\dagger}D\psi\\
 &  & \times\exp\left[N\ln\frac{K}{\lambda}-N+\int dx\left(-2\Phi^{2}+\psi^{\dagger}(\hat{p}+\hat{v}+\hat{a}\gamma_{5}+im+i\frac{(2\pi\rho)^{2}\lambda^{0.5}}{2g}\bar{L}\hat{F}\Phi\cdot\Gamma\hat{F}L)\psi\right)\right]\label{Z2}
\end{eqnarray}
or, integrating over fermions, 
\begin{eqnarray}
\tilde{Z}[v,a,s,p,m]=\int d\lambda D\Phi e^{-\tilde{S}[\lambda,\Phi,v,a,s,p,m]},\label{Z3}
\end{eqnarray}
where 
\begin{eqnarray}
\tilde{S}[\lambda,\Phi,v,a,s,p,m] & = & 2\int dx\,\Phi^{2}+{\rm Tr}\ln\left(\frac{\hat{p}+\hat{v}+\hat{a}\gamma_{5}+im+i\frac{(2\pi\rho)^{2}\lambda^{0.5}}{2g}\bar{L}\hat{F}\Phi\cdot\Gamma\hat{F}L}{\hat{p}+\hat{v}+\hat{a}\gamma_{5}+im}\right)
 +N\ln\frac{K}{\lambda}-N
\label{S}
\end{eqnarray}
and $g^{2}=\frac{(N_{c}^{2}-1)2N_{c}}{2N_{c}-1}$ is a color factor;,
matrices $\Gamma=\{1,\gamma_{5},i\vec{\tau},i\vec{\tau}\gamma_{5}\},$
and we will use notations for the components of the field $\Phi$
in front of them  $\Phi=\{\Phi_{0},\vec{\Phi}\}=\{\sigma,\eta,\vec{\sigma},\vec{\phi}\}$,
with $\Phi^{2}=\Phi_{0}^{2}+\vec{\Phi}^{2}=\sigma^{2}+\eta^{2}+\vec{\sigma}^{2}+\vec{\phi}^{2}$.
Notice that in \re{Z2} in contrast to the NJL model the coupling
constant $\lambda$ is not a parameter of the action but it is defined
by saddle-point condition in the integral \re{Z2}. The partition
function \re{Z2} is invariant under local flavour rotations due
to the gauge links $L$ in the interaction term $\tilde{V}_{\pm,f}[\psi^{\dagger},\psi]$.
However, instead of the explicit violation of the gauge symmetry due
to the zero-mode approximation (\ref{SiDP},\ref{Si}), we have unphysical
dependence of the effective action on the choice of the path in the
gauge link $L$. In our evaluations we used the simplest straight-line
path, though there is no physical reasons why the other choices should
be excluded. In the low-energy region we demonstrated explicitly that
the path dependence drops out.

We will start the calculations of the partition function assuming
for simplicity zero external currents,  and will apply it to evaluation
of the quark condensate $\langle\bar{q}q\rangle$, which is one of
the main manifestations of the SChSB. Technically, it happens due
to the quark-quark interaction term \re{tildeY}, which leads to
the strong attraction in the channels with vacuum (and pion) quantum
numbers. As a consequence,  the vacuum of the theory has a  nonzero
vacuum expectation $\sigma$ of scalar-isoscalar component of meson
fields $\Phi$ and quark condensate $\langle\bar{q}q\rangle$. For
evaluation of the partition function $Z[m]$ it is very convenient
to use the formalism of the effective action \cite{Coleman:1973jx,Jackiw74}
$\Gamma_{eff}[m,\lambda,\Phi]$, which is defined as 
\begin{eqnarray}
 \tilde{Z}[m]=\int d\lambda\tilde{Z}[m,\lambda]=
  \int d\lambda\exp(-\Gamma_{eff}\left[m,\lambda,\Phi\right]),\label{Veff}
\end{eqnarray}
where the field $\Phi$ is the solution of the vacuum equation 
\begin{eqnarray}
\frac{\partial\Gamma_{eff}[m,\lambda,\Phi]}{\partial\Phi}=0,\label{vacuum}
\end{eqnarray}
and implicitly depends on $\lambda$, \textit{i.e.} $\Phi=\Phi(\lambda)$.
In view of~(\ref{vacuum}), the definition~(\ref{Veff}) is equivalent
to the definition via Legendre transformation of partition function,
which is conventionally used in the Quantum Field Theory. The only
nonzero vacuum meson field which might have nonzero vacuum expectation
is a condensate $\Phi=\sigma$. In view of the translation invariance
of  the effective action $\Gamma_{eff}[m,\lambda,\Phi]$, on vacuum
cconfigurations the latter may be replaced with effective potential
$V_{eff}[m,\lambda,\sigma]$.

\begin{figure}[h]
\includegraphics[scale=0.35]{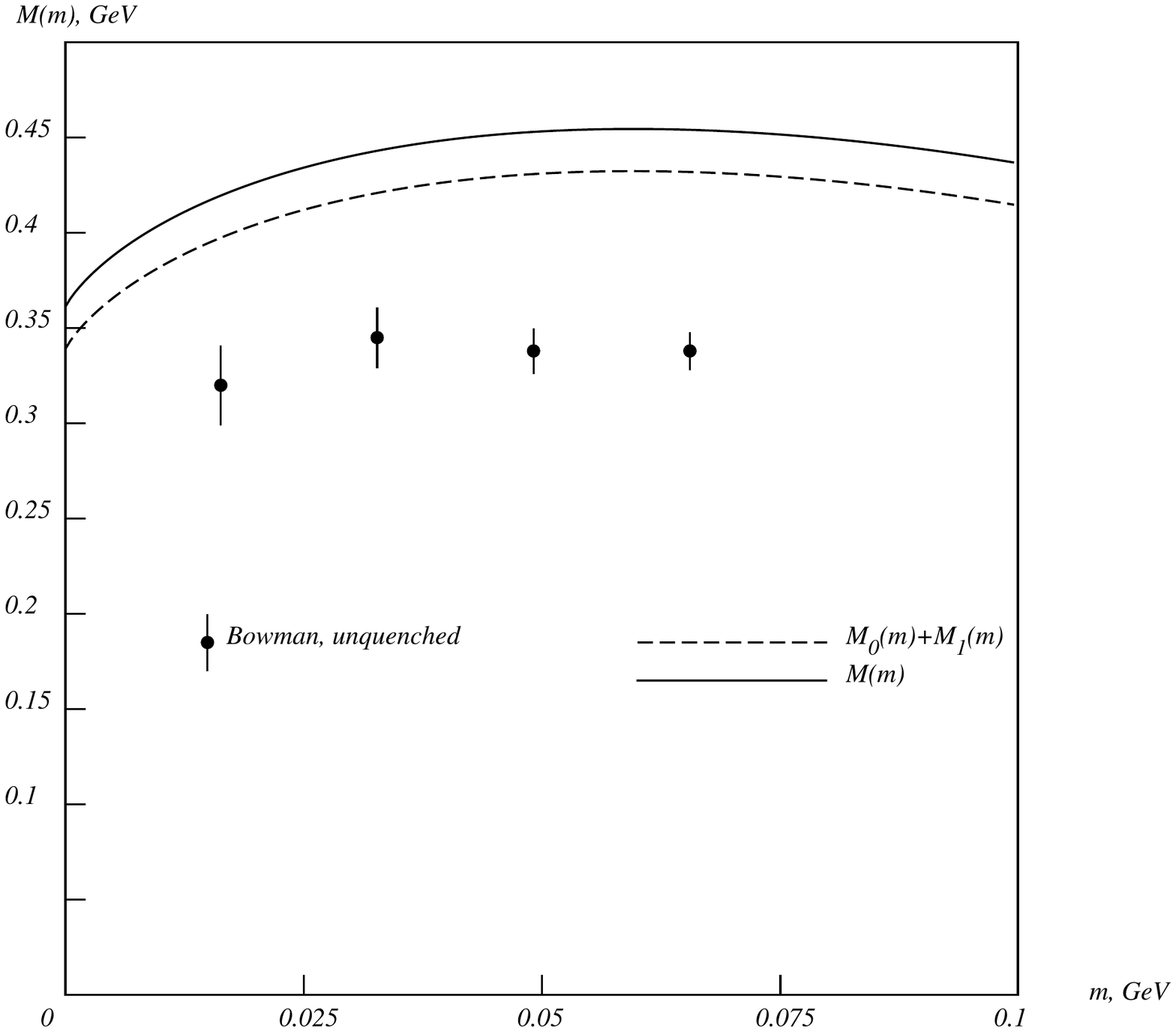}
\includegraphics[scale=0.35]{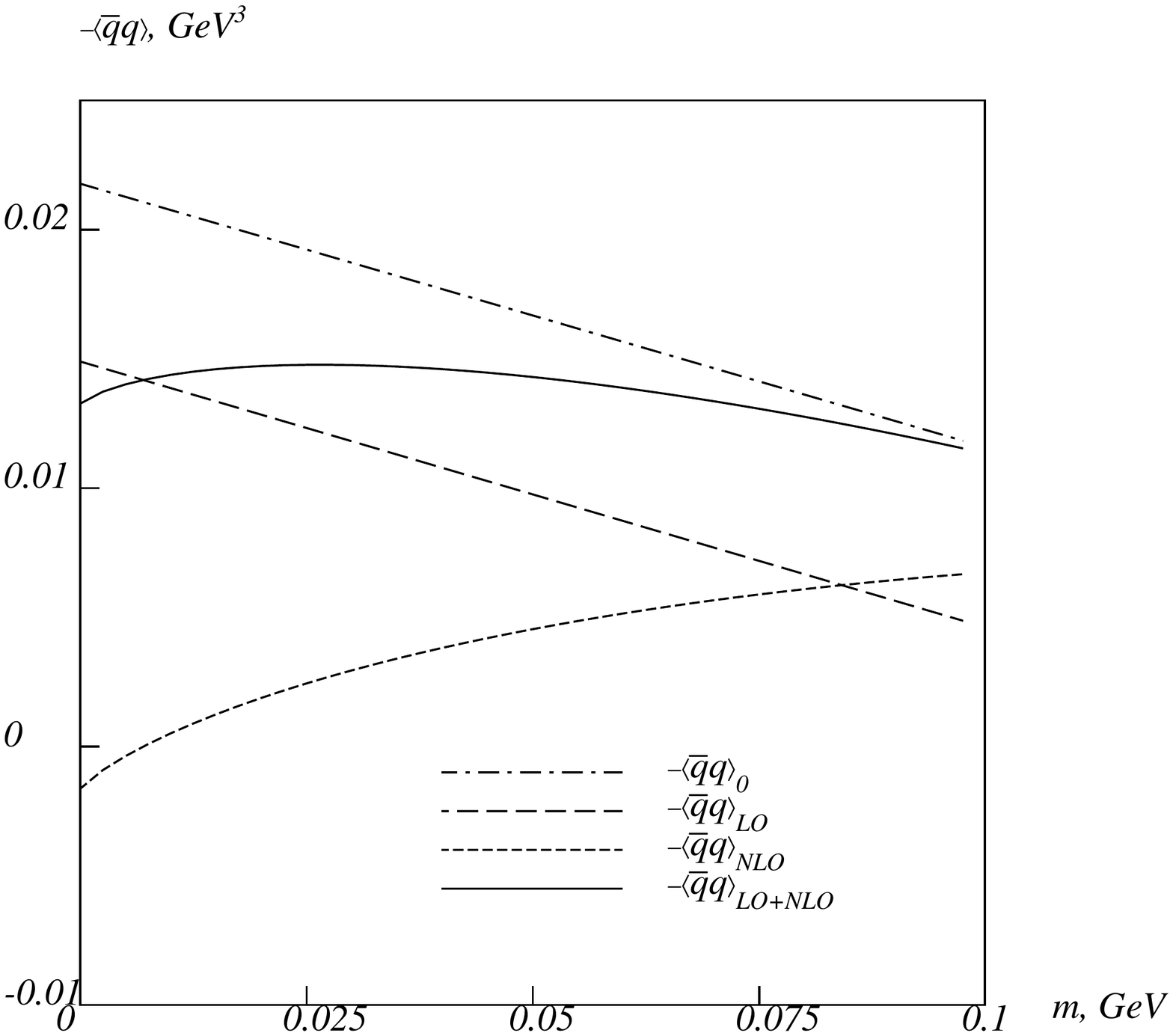}
\caption{\textbf{Left plot:}  The $m$-dependence of the dynamical quark mass
$M$. The solid curve is the exact numerical solution \re{gap1_solution_exact}
of the vacuum Eqs.~\re{vacuum}. The dashed curve is the solution
obtained by the iterations ($1/N_{c}$-expansion) with the same accuracy.
Data points are from~\cite{Bowman:2005vx}. Notice that the scale
of the lattice data is $1.64\,{\rm GeV}$, not $\rho^{-1}\approx0.6\,{\rm GeV}$.
The discrepancy may be attributed to the scales difference. \textbf{Right
plot:}  The $m$-dependence of the quark condensate $-\langle\bar{q}q\rangle$.
The long-dashed curve is the leading-order (LO) result $-<{\bar{q}q}>_{LO}$,
The dot-dashed line represents LO in $1/N_{c}$-expansion result.
The short-dashed curve is the NLO contribution $-<{\bar{q}q}>_{NLO}$,
the solid curve is the total contribution $-<{\bar{q}q}>_{LO+NLO}$. }
\label{fig:1} 
\end{figure}

In absence of quantum (loop) corrections, the effective action just
coincides with the action (\ref{S}). Making expansion of the mesonic
field near its vacuum value, $\Phi\approx\sigma+\Phi'$, and integrating
over the fluctuations, we get for the meson loop correction to the
effective action 
\begin{eqnarray}
\Gamma_{eff}^{mes}[m,\lambda,\sigma]=\frac{1}{2}{\rm Tr}\ln\left(4\delta_{ij}-\frac{1}{\sigma^{2}}{\rm Tr}\frac{M_{q}(p)}{\hat{p}+i(m+M_{q}(p))}\Gamma_{i}\frac{M_{q}(p)}{\hat{p}+i(m+M_{q}(p))}\Gamma_{j}\right).\label{Vmes}
\end{eqnarray}
where $M_{q}(p)$ is the  dynamical quark mass  defined in~(\ref{M}).
Formally, meson loops lead to NLO corrections which are suppressed
as  $\mathcal{O}\left(1/N_{c}\right)$~ in the limit $N_{c}\to\infty$.
We found as result of the calculations These corrections affect all
physical observables, and for example, for the dynamical mass instead
of~(\ref{M}) we get significantly more sophisticated equation whose
solution yields 
\begin{eqnarray}
M(m)=0.36-2.36\,m-\frac{m}{N_{c}}(0.808+4.197\ln m)\label{gap1_solution_exact}
\end{eqnarray}
The essential feature of Eq.~\re{gap1_solution_exact} is the presence
of the chiral logs originated from pion loops, as expected from chiral
perturbation theory~\cite{Gasser:1983yg}. The accuracy of these
solutions is ${\cal O}\left(m^{2},\frac{1}{N_{c}^{2}}\right).$ Left
plot in Fig.\ref{fig:1} represents the $M(m)$-dependence obtained
from Eq. \re{vacuum} with account of $\mathcal{O}(1/N_{c})$~meson
loops corrections. For the sake of comparison, we also plotted the
lattice data from~\cite{Bowman:2005vx}.

The  quark condensate $\langle\bar{q}q\rangle$  can be extracted
 directly from the effective action as 
\begin{eqnarray}
\langle\bar{q}q\rangle=\frac{1}{2}\frac{\partial\Gamma_{eff}}{\partial m}\label{cond}
\end{eqnarray}
Evaluation of (\ref{cond}) gives 
\begin{eqnarray}
 &  & -\langle\bar{q}q\rangle(m)=\left[\left(0.00497-0.0343\,m\right)\,N_{c}+\right.\\
 &  & \left.+\left(0.00168-0.0494\,m-0.0580\,m\,\ln m\right)\right]\left[{\rm GeV}^{3}\right]+{\cal O}\left(m^{2},\frac{1}{N_{c}}\right)\nonumber 
\end{eqnarray}
The $\langle\bar{q}q\rangle(m)$-dependence is depicted in the right
panel of the Fig.~\ref{fig:1}. We can see again that due to the
chiral logarithm the $m$-dependence is not linear, and meson loops
change drastically the $m-$dependence of the quark condensate.

\begin{figure}[h]
 \vskip -3.8cm
\includegraphics[scale=0.35]{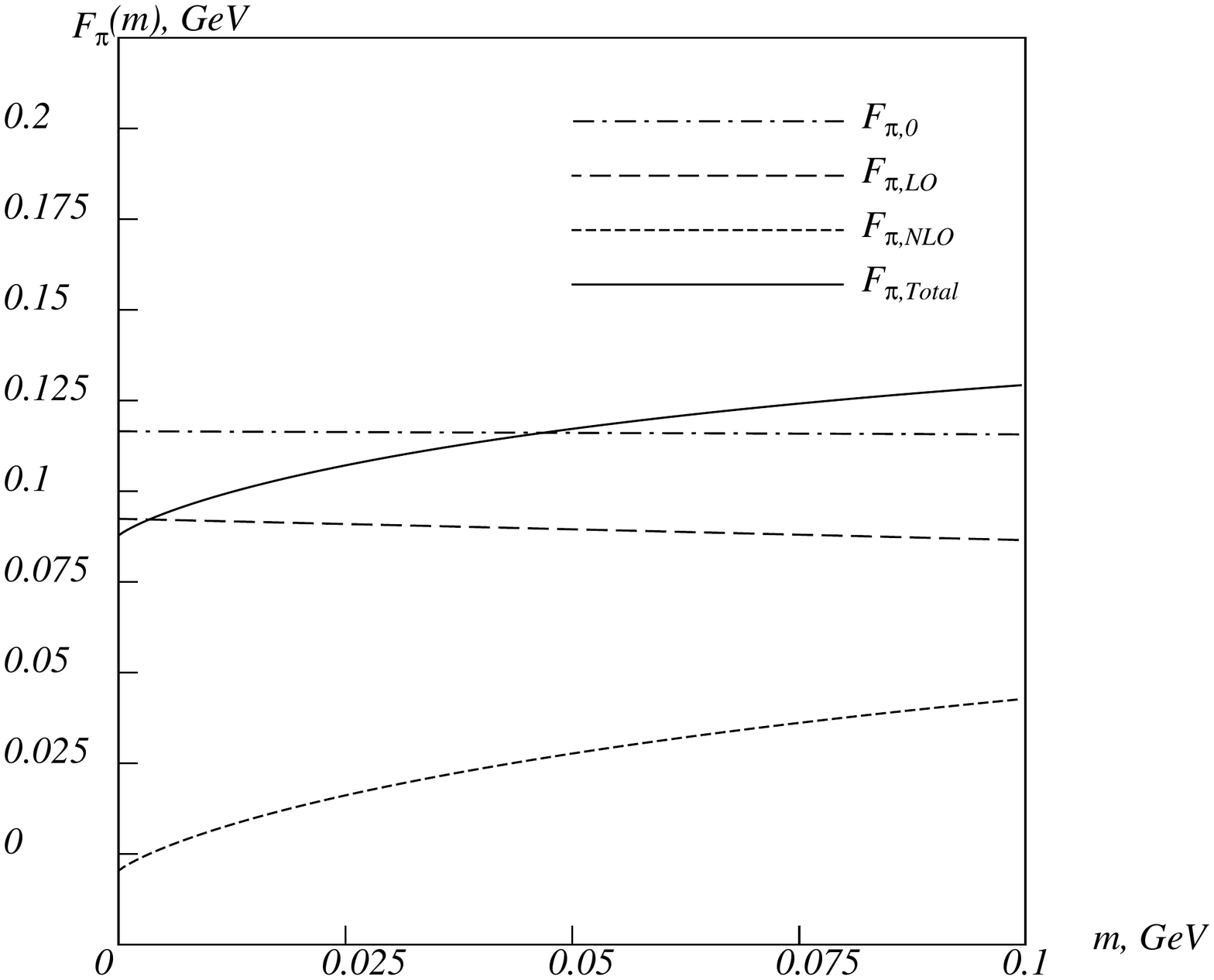}
\includegraphics[scale=0.35]{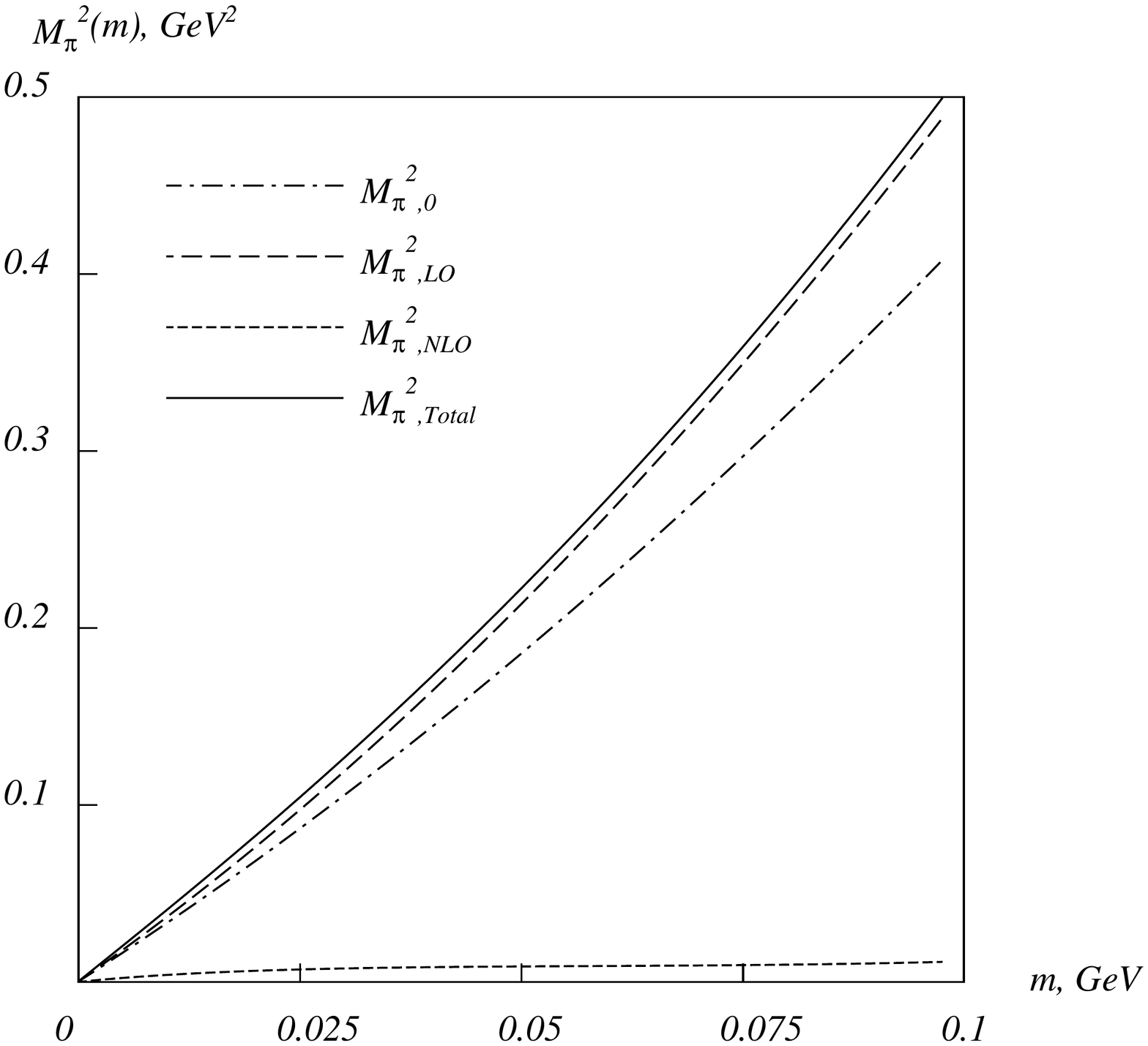}
\caption{\textbf{Left panel:} The $m$-dependence of the pion decay constant
$F_{\pi}$. The long-dashed curve is the LO contribution, the short-dashed
curve is the NLO contribution, the solid curve is the total LO+NLO
contribution. The dot-dashed line represents the leading-order in
$1/N_{c}$-expansion result. \textbf{Right panel:} The $m$-dependence
of the pion mass $M_{\pi}$. The long-dashed curve is the LO contribution,
the short-dashed curve is the NLO contribution, the solid curve is
the total LO+NLO contribution. The dot-dashed line represents the
lowest order in $1/N_{c}$-expansion result. }
\label{fig:Fpi(m)} 
\end{figure}

The extension of this procedure in background of weak external sources
$v,a,s,p$ is straightforward and might be done perturbatively. The
correlators of axial-vector and pseudoscalar currents, which are important
for the Chiral Perturbation Theory ($\chi$PT)~\cite{Gasser:1983yg},
 can be extracted from the Eq.\re{Z3} by taking into account respective
external axial-vector and pseudoscalar sources.  Using the action~(\ref{S})
and taking into account the loop corrections as described earlier,
we may obtain the correlators of axial currents, which allow to extract
the $m$-dependence of the pion decay constant $F_{\pi}$ and pion
mass $M_{\pi}$, 
\begin{eqnarray}
F_{\pi}^{2}=N_{c}\left(\left(2.85-\frac{0.869}{N_{c}}\right)-\left(3.51+\frac{0.815}{N_{c}}\right)m-\frac{44.25}{N_{c}}\,m\,\ln m+{\cal O}\left(m^{2}\right)\right)\cdot10^{-3}\;\left[{\rm GeV}^{2}\right]\label{Res:Fpi}
\end{eqnarray}
where $m$ is given in GeV, the constant in front of ${\cal O}(m^{0})$
contribution is given in ${\rm GeV}^{2}$, and the constant in front
of ${\cal O}(m)$ term is given in ${\rm GeV}$. The $F_{\pi}(m)$-dependence
is shown in the left panel of the Fig.~\ref{fig:Fpi(m)}. For the
sake of comparison, we also plotted the value $F_{\pi,0}$ which we
would get using LO formulae with the mass $M_{0}(m)$. Recall that
the value $F_{\pi}(m=0)=88\,{\rm MeV}$, as well as $<{\bar{q}q(m=0)}>=(255\,MeV)^{3}$,
was used as the input in order to fix the parameters $(\rho,\,R)$.
The comparison between the solid curve and the long-dashed one shows
that the effect of the NLO-corrections grows with $m$ and is about
$40\%$ at $m=0.1\,{\rm GeV}$. Similar calculations for the pion
mass $M_{\pi}(m)$ yield 
\begin{eqnarray}
M_{\pi}^{2}=m\left(\left(3.49+\frac{1.63}{N_{c}}\right)+m\left(15.5+\frac{18.25}{N_{c}}+\frac{13.5577}{N_{c}}\ln m\right)+{\cal O}(m^{2})\right)\label{Res:Mpi}
\end{eqnarray}
in the next-to leading order. The $M_{\pi}(m)$-dependence of the
pion mass is shown in the right panel of Fig. ~\ref{fig:Fpi(m)}.
For the sake of comparison, we also plotted the value $M_{\pi,0}$.
which one would get using LO formulae with the mass $M_{0}(m)$. Altogether,
for this observables the NLO-corrections turn out to be small. 
 These results are essential and allow for the extraction of the low-energy
constants of Chiral Perturbation Theory (ChPT), matching the expected
$m$-dependence in that theory with $m$-dependence found in Instanton
Liquid Model. 
. According to~\cite{Gasser:1983yg}, the low-energy constants $\bar{l}_{i}$
of the chiral lagrangian appear in some of the ${\cal O}(m)$-corrections
to physical quantities, e.g. 
\begin{eqnarray}
M_{\pi}^{2}=m_{\pi}^{2}\left(1-\frac{m_{\pi}^{2}}{32\pi^{2}F^{2}}\bar{l}_{3}+{\cal O}(m_{\pi}^{4})\right),\,\,F_{\pi}^{2}=F^{2}\left(1+\frac{m_{\pi}^{2}}{8\pi^{2}F^{2}}\bar{l}_{4}+{\cal O}(m_{\pi}^{4})\right),
\end{eqnarray}
where $M_{\pi},F_{\pi}$ are the pion mass and decay constants, $m_{\pi}^{2}=2\,m\,B$
and $B,F$ are the phenomenological parameters of the chiral lagrangian.
Using our results (\ref{Res:Fpi}, \ref{Res:Mpi}), we can obtain
\begin{eqnarray}
 &  & F^{2}=0.00284777N_{c}-0.000868917+{\cal O}\left(\frac{1}{N_{c}}\right)%
,\,\,\,B=1.7467+\frac{0.8183}{N_{c}}+{\cal O}\left(\frac{1}{N_{c}^{2}}\right)\label{F2}\\
 &  & \bar{l}_{3}=0.0738267-1.14251\,N_{c}-0.999\,\ln\left(\frac{m}{{\rm GeV}}\right)+{\cal O}\left(\frac{1}{N_{c}}\right)\label{L_3}\\
 &  & \bar{l}_{4}=-0.0793814\,N_{c}+0.0187608-1.000\,\ln\left(\frac{m}{{\rm GeV}}\right)+{\cal O}\left(\frac{1}{N_{c}}\right)\label{L_4}
\end{eqnarray}
which gives numerically
\begin{eqnarray}
F=88\,{\rm MeV},\;B=2.019\,{\rm GeV},\;\bar{l}_{3}=1.84,\;\bar{l}_{4}=4.98\label{l3l4:values}
\end{eqnarray}
for current mass $m=0.0055\,\,{\rm GeV}$; the corresponding values
of the pion mass and decay constant are given by $M_{\pi}=0.142\,\,{\rm GeV}$,
$F_{\pi}=0.0937\,\,{\rm GeV}$. We need to mention that the values
of $F,-\langle{\bar{q}q(m=0)}\rangle=-F^{2}B$ in (\ref{l3l4:values})
were taken as input when we fixed the parameters $\bar{\rho},\,R$.
Our results for $\bar{l}_{3},\bar{l}_{4}$  are in fair agreement
with the phenomenological estimates~\cite{Leutwyler:2006qq,Leutwyler:2007ae}
and lattice predictions~\cite{Aubin:2004fs,DelDebbio:2006cn} given
in Table~\ref{Table:l_i}. 
\begin{table}[h]
\begin{tabular}{|c|c|c|c|c|c|}
\hline 
{} & {ChPT} & {MILC} & {Del Debbio} & {ETM} & {Our}\tabularnewline
 & \cite{Colangelo:2001df,Gasser:1983yg,Leutwyler:2007ae} & \cite{Bernard:2006zp} & {\textit{et. al.}~\cite{DelDebbio:2006cn}} & \cite{Boucaud:2007uk} & {prediction}\tabularnewline
\hline 
{$\bar{l}_{3}$} & {$2.9\pm2.4$} & {$0.6\pm1.2$} & {$3.0\pm0.5$} & {$3.62\pm0.12$} & {$1.84$}\tabularnewline
{$\bar{l}_{4}$} & {$4.4\pm0.2$} & {$3.9\pm0.5$} & {---} & {$4.52\pm0.06$} & {$4.98$}\tabularnewline
\hline 
\end{tabular}\caption{\label{Table:l_i}Estimates and predictions of the low-energy constants
$\bar{l}_{3},\,\bar{l}_{4}$. The first column contains phenomenological
estimates, the next three columns are lattice results from different
collaborations, the last column contains our results. The first four
columns of the table are taken from~\cite{Leutwyler:2007ae}.}
\end{table}

Notice that in (\ref{L_3} - \ref{l3l4:values}) we keep ${\cal O}(N_{c},N_{c}^{0})$-terms
and drop NNLO terms ${\cal O}(1/N_{c})$ in agreement with our general
framework. Without such expansion we would get 
\begin{eqnarray}
\bar{l}_{3}=0.28, &  & \bar{l}_{4}=4.28.
\end{eqnarray}
Also, other similar  ChPT low energy parameters were successfully calculated within ILM~\cite{Kim:2004hd,Goeke:2007nc,Goeke:2010hm}. So, ILM is reliable nonperturbative QCD tool for low energy light quarks processes.

\section{Heavy and light quarks systems in ILM}

\label{Qq} As we found above, the strength of instanton-light quark
interaction  is fairly large and induces strong interaction between
light quarks, leading to formation of almost massless pions and nonzero
quark condensate, in accordance with expectations of SBChS. We have
seen that the physics in this domain is closely related to the low-energy
behaviour of the light quark propagator, which is dominated by its
zero mode. On the other hand, heavy quarks dynamics is controlled
mainly by their heavy masses $M_{Q}$, and we expect rather weak instanton-heavy
quark interaction. Nevertheless, we expect essential influence of
instantons for the heavy-light quarks systems.

\subsection{The strength of heavy-quark-instanton interaction}

\label{QI} First, we simplify the problem by taking an infinitely
heavy quark, which interacts only through the fourth components of
instantons $A_{4}$. Hereafter, we follow the definitions given in
Ref.~\cite{Diakonov:1989un}, i.e. $\theta$ is inverse of differentiation
operator $\theta^{-1}=d/dt$ and $\langle t|\theta|t'\rangle=\theta(t-t')$
is a step-function. For the sake of convenience, we also redefine
the gluonic field as $A\equiv iA_{4}$. Accordingly,  the heavy quark
$Q$ and antiquark $\bar{Q}$ Lagrangians can be expressed as 
\begin{eqnarray}
L_{Q}=Q^{+}(\theta^{-1}-gA+...)Q,\,\,\,L_{\bar{Q}}=\bar{Q}^{+}(\theta^{-1}-g\bar{A}+...)\bar{Q},
\end{eqnarray}
where the dots denote the next order in the inverse of heavy quark
mass terms. In terms of SU($N_{c}$) generators, the fields $A$ and
$\bar{A}$ are given as $A=A_{a}\lambda_{a}/2$ and $\bar{A}_{a}=A_{a}\bar{\lambda}_{a}/2$,
where $\bar{\lambda}_{a}=-\lambda_{a}^{{\rm T}}$ (here `T' means
the transposition). In our calculations we may neglect  the virtual
processes $Q\rightarrow QQ\bar{Q}$ corresponding to the heavy quark
loops, which effectively means the heavy quark determinant equals
to unity. The functional space of heavy quarks $Q$ is not overlapping
with the functional space of heavy antiquarks $\bar{Q}$ and, consequently,
the total functional space is a direct product of $Q$ and $\bar{Q}$
spaces.

After averaging over instantons collective coordinates, the propagator
of the infinitely heavy quark $Q$  in ILM is given by 
\begin{eqnarray}
{w}=\!\int\!D\xi\frac{1}{\theta^{-1}-gA(\xi)}.\label{wQ}
\end{eqnarray}
In order to integrate over instanton collective coordinates $\xi_{i}$
one can now apply Pobylitsa's equation from Ref.~\cite{Diakonov:1989un}
and follow their solution in the lowest order in packing parameter
$\kappa$, 
\begin{eqnarray}
w^{-1}=\theta^{-1}-\sum_{i}^{N}\int d\xi_{i}\theta^{-1}\left(\frac{1}{\theta^{-1}-gA_{i}(\xi_{i})}-\theta\right)\theta^{-1}+O(\kappa^{2}).\label{wQ-1}
\end{eqnarray}
The calculations of the matrix element of the second term in the right-hand
side of Eq.~(\ref{wQ-1}) provides the direct ILM heavy quark mass
shift $\Delta M_{Q}^{{\rm dir}}$ with  corresponding order ${\cal O}(\kappa)$.
We note that the direct instanton contribution to the heavy quark mass 
$\Delta M_{Q}^{{\rm dir}}$ was calculated first
in Ref.~\cite{Diakonov:1989un}. Numerically, one can estimate 
\begin{eqnarray}
\Delta M_{Q}^{{\rm dir}}\simeq70\,{\rm MeV}\sim\kappa\rho^{-1}.
\label{DeltaMQ}
\end{eqnarray}
This value controls the strength of heavy-quark-instanton interaction,
which is much weaker than that of the light quarks, in agreement with
expectations based on heavy quark mass limit.

\subsection{Heavy and light quarks interactions in ILM}

\label{Qqinteractions}

It is obvious that the light quarks affect the integration in~\re{wQ}
via the contributions of the light quark determinant. For our aim
it is useful to take into account light quarks sources, Eq.~\re{part-func},
 and extend the Eq.~\re{wQ} in the form 
\begin{eqnarray}
 & w=\int\prod_{f}D\psi_{f}D\psi_{f}^{\dagger}\exp\int\sum_{f}\left(\psi_{f}^{\dagger}(\hat{p}\,+\,im_{f})\psi_{f}\right)\prod_{\pm}^{N_{\pm}}H_{\pm}[\psi^{\dagger},\psi]w[\psi,\psi^{\dagger}],\label{w}
 \\\label{H}
 & H_{\pm}[\psi^{\dagger},\psi]=
 \int d\xi_{\pm}\prod_{f}V_{\pm,f}[\psi^{\dagger},\psi,\xi_{\pm}],
\\
\label{wpsi}
 & w[\psi,\psi^{\dagger}]=\left\{ \prod_{\pm}^{N_{\pm}}H_{\pm}[\psi^{\dagger},\psi]\right\} ^{-1}\!\!\!\int\prod_{\pm}^{N_{\pm}}d\xi_{\pm}\left\{ \prod_{\pm}^{N_{\pm}}V_{\pm,f}[\psi^{\dagger},\psi,\xi_{\pm}]\right\} \left(\theta^{-1}-\sum_{i}A_{i}\right)^{-1}.
\end{eqnarray}
The measure of the integration over $\xi_{\pm}$ in the Eq.~(\ref{wpsi})
with and without light quark factor $\prod_{f}V_{\pm,f}[\psi^{\dagger},\psi]$
has the same structure as a product of independent integrations over
the instanton collective coordinates $\xi_{\pm}$. Then, we may extend
the derivation of the Pobylitca equations~\cite{Diakonov:1989un,Pobylitsa:1989uq}
and solve them in the dilute approximation as 
\begin{eqnarray}
w^{-1}[\psi,\psi^{\dagger}]=\theta^{-1}-\frac{N}{2}\sum_{\pm}\frac{1}{H_{\pm}[\psi^{\dagger},\psi]}\Delta_{Qq,\pm}[\psi^{\dagger},\psi]+O(\kappa^{2}),\label{w-1}
\end{eqnarray}
where, defining the heavy quark propagator in the single (anti)instanton
field as $w_{\pm}=(\theta^{-1}-A_{\pm})^{-1}$, we have 
\begin{eqnarray}
\Delta_{Qq,\pm}[\psi^{\dagger},\psi]=\int d\xi_{\pm}\prod_{f}V_{\pm,f}[\psi^{\dagger},\psi,\xi_{\pm}]\theta^{-1}(w_{\pm}[\xi_{\pm}]-\theta)\theta^{-1}.\label{Delta}
\end{eqnarray}
The last expression represents the interactions of heavy $Q$ and
$N_{f}$ light quarks $q$ generated by instantons as (see Fig.~\ref{figQq})
\begin{eqnarray}
S_{Qq}=-\lambda\sum_{\pm}\int Q^{\dagger}\Delta_{Qq,\pm}[\psi^{\dagger},\psi]Q.
\end{eqnarray}
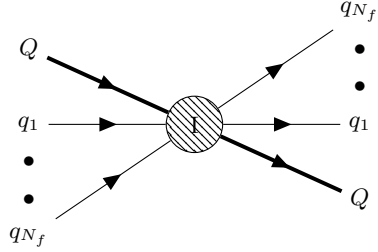
\begin{figure}[!ht]
\center \tikz{\begin{feynman} \vertex (i1){$Q$}; \vertex(q1)
[below=1cm of i1]{$q_{1}$}; \vertex(m1) [below=0.5cm of
q1]{$\bullet$}; \vertex(n1) [below=0.5cm of m1]{$\bullet$};
\vertex(o1) [below=0.5cm of n1]{$q_{N_{f}}$}; \vertex(c)[blob]
[right=2.2cm of q1]{${\rm I}$}; \vertex(q2) [right=2.2cm
of c]{$q_{1}$}; \vertex(i2) [below=1cm of q2]{$Q$}; \vertex(n2)
[above=0.5cm of q2]{$\bullet$}; \vertex(m2) [above=0.5cm
of n2]{$\bullet$}; \vertex(o2) [above=0.5cm of m2]{$q_{N_{f}}$};
\diagram*{ (i1) --[fermion,line width=1.5pt] (c) --[fermion,line
width=1.5pt] (i2), (q1) --[fermion] (c) --[fermion] (q2),
(o1) --[fermion] (c) --[fermion](o2); };\end{feynman}}
\caption{Instanton generated heavy $Q$ and $N_{f}$ light $q$ quarks interaction.}
\label{figQq} 
\end{figure}

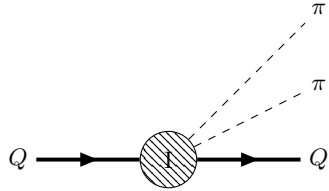
\begin{figure}[!ht]
\center \tikz{\begin{feynman} \vertex (i1){$Q$}; \vertex(i2)[blob]
[right=2cm of i1]{${\rm I}$}; \vertex(i3) [right=2cm of
i2]{$Q$}; \vertex(o1) [above=1cm of i3]{$\pi$}; \vertex(o2)
[ 
above=2cm of i3]{$\pi$}; \diagram*{ (i1) --[fermion,line
width=1.5pt] (i2) --[fermion,line width=1.5pt] (i3), (i2) --[scalar]
(o1), (i2) --[scalar] (o2); };\end{feynman}} \caption{Instanton generated emission of 2 pions by heavy $Q$ quark.}
\label{figQpi} 
\end{figure}

It is straightforward to see that without light quarks factors $V_{\pm,f}[\psi^{\dagger},\psi,\xi_{\pm}]$,
the Eq.~\re{Delta} reduces to the second term of Eq.~(\ref{wQ-1}),
which gives the direct ILM contribution to the heavy quark mass $\Delta M_{Q}^{{\rm dir}}$.
It is obvious that the light quark pairs are produced in color singlet
or color octet states. Let us consider $N_{f}=2$ colorless light
quark pairs. These states are represented by mesons, and we'll focus
on the lightest of them, pions, as shown in Fig.\,\ref{figQpi}.

Technically, it can be done by bosonization method (see subsection~\ref{mesons}
for details). The corresponding amplitude of the emission of pions
will have a form~\cite{Musakhanov2018} 
\begin{eqnarray}
A_{Q\pi}=-F_{\pi Q}^{2}\int\frac{d^{4}x}{(2\pi)^{4}}e^{-ipx}{\rm tr}_{f}\partial_{\mu}U(x)\partial_{\mu}U^{+}(x)\Delta M_{Q}^{{\rm dir}}R^{4}F_{Q}(p),\,\,F_{Q}(p)=\frac{J_{0}(|\vec{p}|\rho)}{J_{0}(0)}.\label{AQpions}
\end{eqnarray}
where $J_{0}(|\vec{p}|\rho)$ is a Bessel function, and 
\begin{eqnarray}
F_{\pi Q}^{2}=2N_{c}\int\frac{d^{4}p}{(2\pi)^{4}}\frac{M^{2}(p)p^{2}}{(p^{2}+M^{2}(p))^{3}},\label{FpiQ}
\end{eqnarray}
%
where $M(p)$ is the dynamical light quark mass. At the values of ILM
parameters $\rho=1/3\,{\rm fm},\,\,\,R=1\,{\rm fm}$
one can get $F_{\pi Q}=0.6 F_{\pi}$. Here  the pion decay constant
value in LO accordingly Eq.~(\ref{F2}) is given by $F_\pi=92.4\,{\rm MeV}$.
The integration over $x$ in Eq.~(\ref{AQpions}) gives the $\delta$-function
which reflects the energy-momentum conservation.

\subsection{Heavy quarkonium and light quarks interactions in ILM}

\label{QQq}

Now we will consider in ILM the $Q\bar{Q}$ correlator with the account
of light quarks, which is defined as 
\begin{eqnarray}\nonumber
 &  & \langle0|T(Q^{\dagger}(t_{2}',\vec{x}_{2})\bar{Q}^{\dagger}(t_{1}',\vec{x}_{1}))\bar{Q}(t_{2},\vec{x}_{2})Q(t_{1},\vec{x}_{1})|0\rangle=\int D\psi_{f}D\psi_{f}^{\dagger}\left\{ \prod_{\pm}^{N_{\pm}}H_{\pm}[\psi^{\dagger},\psi]\right\} 
 \\
 &  & \times\exp\left(\int\sum_{f}\psi_{f}^{\dagger}(\hat{p}+im_{f})\psi_{f}\right)
\langle t_{1}',\vec{x}_{1};t_{2}',\vec{x}_{2}|W[\psi,\psi^{\dagger}]|t_{1},\vec{x}_{1};t_{2},\vec{x}_{2}\rangle 
 \label{QQcorrelator}
\end{eqnarray}
where 
\begin{eqnarray}\nonumber
 &  & \langle t_{1}',\vec{x}_{1};t_{2}',\vec{x}_{2}|W[\psi,\psi^{\dagger}]|t_{1},\vec{x}_{1};t_{2},\vec{x}_{2}\rangle=\left\{ \prod_{\pm}^{N_{\pm}}H_{\pm}[\psi^{\dagger},\psi]\right\} ^{-1}\int D\xi\left\{ \prod_{\pm}^{N_{\pm}}\prod_{f}V_{\pm,f}[\psi_{f}^{\dagger},\psi_{f},\xi_{\pm}]\right\} 
 \\\label{WQQq}
 &  & \times\langle t_{1}',\vec{x}_{1}|\left(\theta^{-1}-\sum_{i}A_{i}^{(1)}\right)^{-1}|t_{1},\vec{x}_{1}\rangle\otimes\langle t_{2}',\vec{x}_{2}|\left(\theta^{-1}-\sum_{i}\bar{A}_{i}^{(2)}\right)^{-1}|t_{2},\vec{x}_{2}\rangle 
 \\\nonumber {\rm and}
 &  & \langle t_{1}',\vec{x}_{1}|\left(\theta^{-1}-\sum_{i}A_{i}^{(1)}\right)^{-1}|t_{1},\vec{x}_{1}\rangle\otimes\langle t_{2}',\vec{x}_{2}|\left(\theta^{-1}-\sum_{i}\bar{A}_{i}^{(2)}\right)^{-1}|t_{2},\vec{x}_{2}\rangle
 \\
 & = &\left[T\exp\left(ig\int_{t_{1}'}^{t_{1}}d\tau_{1}\sum_{i}A_{i,4}(\xi,\vec{x}_{1},\tau_{1})\right)\right]\left[T\exp\left(ig\int_{t_{2}'}^{t_{2}}d\tau_{2}\sum_{i}\bar{A}_{i,4}(\xi,\vec{x}_{2},\tau_{2})\right)\right],
 \label{WQQ}
\end{eqnarray}
where $\otimes$ is a tensor product and the fields $A_{i}^{(1)}$ and $\bar{A}_{i}^{(2)}$ are the projections
of the instanton fields onto the lines $L_{1}$ and $L_{2}$ corresponding
to the heavy quark $Q$ and the heavy antiquark $\bar{Q}$, respectively.
\begin{figure}[h]
\begin{centering}
\includegraphics[scale=0.7]{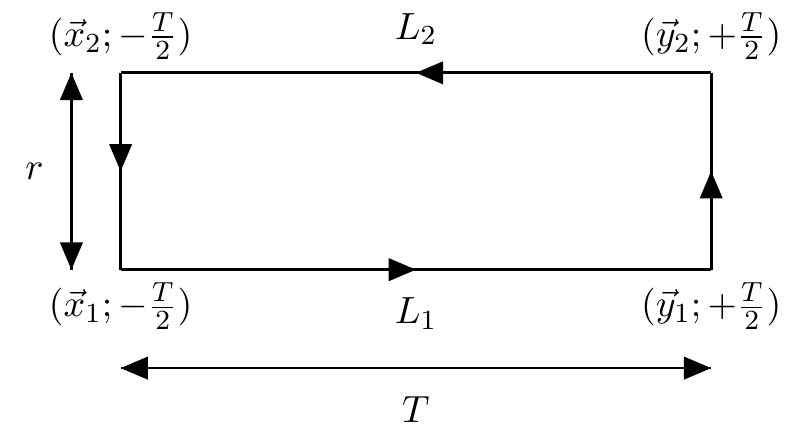}
\par\end{centering}
\caption{The Wilson loop, corresponding to $Q\bar{Q}$ correlator.}
\label{W-loop} 
\end{figure}
Under the same argumentation as before (see Eq.~(\ref{w-1})), one
may extend Pobyitca's Eq.~\cite{Pobylitsa:1989uq} and,  neglecting
${\cal O}(\kappa^{2})$ terms, get the solution of extended equation
\begin{eqnarray}\nonumber
 &  & W^{-1}[\psi,\psi^{\dagger}]=w^{(1)-1}[\psi,\psi^{\dagger}]\otimes\bar{w}^{(2)-1}[\psi,\psi^{\dagger}]-\frac{N}{2}\sum_{\pm}\frac{1}{H_{\pm}[\psi^{\dagger},\psi]}\int d\xi_{\pm}\prod_{f}V_{\pm,f}[\psi_{f}^{\dagger},\psi_{f},\xi_{\pm}]\\
 &  & \times\int d\xi_{\pm}\prod_{f}V_{\pm,f}[\psi_{f}^{\dagger},\psi_{f},\xi_{\pm}]\left[\left(\theta^{-1}(w_{\pm}^{(1)}-\theta)\theta^{-1}\right)\otimes\left(\theta^{-1}(\bar{w}_{\pm}^{(2)}-\theta)\theta^{-1}\right)\right].
 \label{WQQq-1}
\end{eqnarray}
The Eq.~(\ref{WQQq-1}) describes
a heavy quark-antiquark $Q\bar{Q}$ pair interacting with $N_{f}$
light quarks (see Fig.~\ref{figQQq}). We see from the structure
of the first and second terms that they interfere strongly, and light
quarks are emitted only from the region where heavy quarks $Q$ and
$\bar{Q}$ stay inside the same instanton. Moreover, for colorless
$Q\bar{Q}$ system with relative distance $r=0$ this emission must
disappear. It can be understood from the consideration of the instanton
contribution to the total energy  $\Delta E^{{\rm dir}}(r)$ of $Q\bar{Q}$
colorless system extracted from the asymptotics ($T\rightarrow\infty$)
of Eq.~\re{WQQq} by neglecting the light quarks factors in it (
this quantity can be extracted directly from Eq.~\re{WQQq-1}).
We have $\Delta E^{{\rm dir}}(r)=2\Delta M_{Q}^{{\rm dir}}+V^{{\rm dir}}(r)$.
It is clear that $\Delta E^{{\rm dir}}(\infty)=2\Delta M_{Q}^{{\rm dir}}$,
while $\Delta E^{{\rm dir}}(0)=0$, since at zero separation between
$Q$ and $\bar{Q}$ in colorless state their color multipole momentas
become zero and there are no interaction with instantons. 
\begin{figure}[!ht]
\center \tikz{\begin{feynman} \vertex (i1){$Q$}; \vertex(q1)
[below=0.7cm of i1]{$q_{1}$}; \vertex(m1) [below=0.4cm of
q1]{$\bullet$}; \vertex(n1) [below=0.4cm of m1]{$\bullet$};
\vertex(o1) [below=0.4cm of n1]{$q_{N_{f}}$}; \vertex(j1)
[below=0.7cm of o1]{$\bar{Q}$}; \vertex(c1) [right=2.2cm
of i1]{${\rm }$}; \vertex(c)[blob] [below=1.3cm of c1]{${\rm I}$};
\vertex(i2) [right=2.2cm of c1]{$\bar{Q}$}; \vertex(o2) [below=0.7cm
of i2]{$q_{N_{f}}$}; \vertex(n2) [below=0.4cm of o2]{$\bullet$};
\vertex(m2) [below=0.4cm of n2]{$\bullet$}; \vertex(q2) [below=0.4cm
of m2]{$q_{1}$}; \vertex(j2) [below=0.7cm of q2]{$Q$};
\diagram*{ (i1) --[fermion,line width=1.5pt] (c) --[fermion,line
width=1.5pt] (j2), (j1) --[anti fermion,line width=1.5pt] (c)
--[anti fermion,line width=1.5pt] (i2), (q1) --[fermion]
(c) --[fermion] (q2), (o1) --[fermion] (c) --[fermion](o2);
};\end{feynman}} \caption{Instanton generated heavy $Q\bar{Q}$ heavy quarks and $N_{f}$ $q$
light quarks interaction.}
\label{figQQq} 
\end{figure}
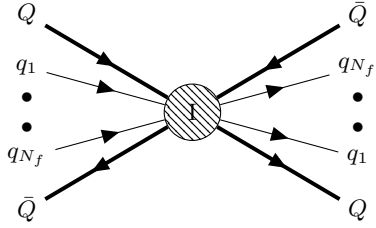


So, in the ILM without light quarks the Eqs.~(\ref{WQQq}, \ref{WQQq-1})
provide the direct instanton contribution to the $Q\bar{Q}$ potential
as we discussed above 
\begin{eqnarray}
V_{{\rm dir}}(r)=\frac{N}{2VN_{c}}\sum_{\pm}\int d^{3}z_{\pm}{\rm tr}_{c}\left[1-P\exp\left(i\int_{L_{1}}dtA_{\pm,4}\right)\right]\left[1-P\exp\left(-i\int_{L_{2}}dtA_{\pm,4}\right)\right].\label{Vdir}
\end{eqnarray}
We have to note that similar quantity $V_{{\rm DPP}}(r)$ was originally
derived in~\cite{Diakonov:1989un}, which differs from Eq.~\re{Vdir}
as 
\begin{eqnarray}
V_{{\rm DPP}}(r)\equiv\Delta E^{{\rm dir}}(r)=V_{{\rm dir}}(r)+2\Delta M_{Q}^{{\rm dir}}
\end{eqnarray}
At small ($x\ll\rho$) and large distances ($x\gg\rho$), it can be
evaluated analytically and one has the following form of the potential
\begin{eqnarray}
 &  & \Delta E^{{\rm dir}}(r<<\rho)\simeq\frac{4\pi^{4}\kappa}{3N_{c}}\left[\frac{\pi}{16}-J_{1}(2\pi)\right]\frac{r^{2}}{\rho}+...\\
 &  & \Delta E^{{\rm dir}}(r>>\rho)\simeq2\Delta M_{Q}^{{\rm dir}}-\frac{2\pi^{3}\kappa}{N_{c}r}+....
\end{eqnarray}
in terms of the Bessel functions $J_{n}$. An average size of charmonium
is comparable with the instanton size $r_{c}\sim\rho$ while for botomonium
the relation $r_{b}<\rho$ is valid. Consequently, one may expect
that $r^{2}$-approximation will work better in the botomonium case
in comparison with the charmonium case. It is well know that $r^{2}$-approximation
corresponds to the dipole approximation in the multi-pole expansion.

\begin{figure}[!ht]
\center \tikz{\begin{feynman} \vertex (i1){$Q$}; \vertex
(o1) [below=3cm of i1]{$\bar{Q}$}; \vertex(c)[blob] [below
right=2.2cm of i1]{${\rm I}$}; \vertex(i3) [below right=2.2cm
of c]{$Q$}; \vertex(o2) [above right=2.2cm of c]{$\bar{Q}$};
\vertex(p1) [above=1cm of i3]{$\pi$}; \vertex(p2) [below=1cm
of o2]{$\pi$}; \diagram*{ (i1) --[fermion,line width=1.5pt]
(c) --[fermion,line width=1.5pt] (i3), (o1) --[anti fermion,line
width=1.5pt] (c) --[anti fermion,line width=1.5pt](o2), (c)--[scalar](p1),
(c)--[scalar](p2); };\end{feynman}} \caption{Instanton generated emission of pions by heavy $Q\bar{Q}$ quark system.\label{figQQpion} }
\end{figure}
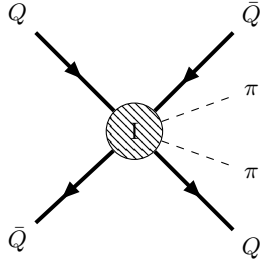

The interaction term in the Eq.~(\ref{WQQq-1}) has a part corresponding
to the colorless state of light quarks. From this part we can calculate
the amplitude $A_{QQ\pi}$ of the process $(Q\bar{Q})_{n'}\rightarrow(Q\bar{Q})_{n}\pi\pi$
(see Fig.~\ref{figQQpion}), corresponding to the $N_{f}=2$ case
\begin{eqnarray}
 &  & A_{QQ\pi}=F_{\pi Q}^{2}\int d^{4}z\,{\rm tr}_{f}\partial_{\mu}U(z)\partial_{\mu}U^{+}(z)\exp(i(\vec{p}'-\vec{p})\vec{z})\\
 &  & \times\int d^{3}yd^{3}r\langle n|\vec{r}\rangle\exp(-i\vec{p}\vec{y})\frac{1}{N_{c}}{\rm tr}_{c}\left\{ 1-P\exp\left(i\int_{-\infty}^{\infty}d\tau_{1}A_{\pm,4}(\vec{y}+\vec{r}/2,\tau_{1})\right)\right.\\
 &  & \left.\times\,P\exp\left(-i\int_{-\infty}^{\infty}d\tau_{2}A_{\pm,4}(\vec{y}-\vec{r}/2,\tau_{2})\right)\right\} \exp(i\vec{p}^{\,\prime}\vec{y})\langle\vec{r}|n'\rangle,\label{AQQpions3}
\end{eqnarray}
where $\vec{y}=\vec{x}-\vec{z}$, and the positions of $Q$ and $\bar{Q}$
are taken as $\vec{x}_{1}=\vec{x}+\vec{r}/2$ and $\vec{x}_{2}=\vec{x}-\vec{r}/2$.

In Eq.\,(\ref{AQQpions3}), $\exp(i\vec{p}^{\,\prime}\vec{x})|n'\rangle$
and $\exp(i\vec{p}\vec{x})|n\rangle$ are the initial and final states
of $Q\bar{Q}$ system with the corresponding total momentums $\vec{p}^{\,\prime}=(\vec{p}_{1}^{\,\prime}+\vec{p}_{2}^{\,\prime})$
and $\vec{p}=(\vec{p}_{1}+\vec{p}_{2})$, respectively. They are solutions
of the Schrodinger equation with the Hamiltonian 
\begin{eqnarray}
H_{0}=T+V,\qquad T=\frac{{\vec{p}_{1}}^{\,2}}{2m_{Q}}+\frac{{\vec{p}_{2}}^{\,2}}{2m_{Q}},\label{HQQ}
\end{eqnarray}
where $V$ is the $Q\bar{Q}$ potential in the ILM, accompanied by
phenomenological confining potential. We would like to mention that
one of the most important components in $V$ is the perturbative QCD
contribution. The matrix element between the heavy quarkonium $Q\bar{Q}$
states in the amplitude~(\ref{AQQpions3}) has the factor $F(\vec{r},\vec{p}^{\,\prime}-\vec{p})$,
which is given by 
\begin{eqnarray}
 &  & F(\vec{r},\vec{p}^{\,\prime}-\vec{p})=\frac{1}{N_{c}}\int d^{3}y\exp(i(\vec{p}^{\,\prime}-\vec{p})\vec{y})\label{structure}\\
 &  & \times{\rm tr}_{c}\left\{ 1-P\exp\left(i\int_{-\infty}^{\infty}d\tau_{1}A_{\pm,4}(\vec{y}+\vec{r}/2,\tau_{1})\right)%
P\exp\left(-i\int_{-\infty}^{\infty}d\tau_{2}A_{\pm,4}(\vec{y}-\vec{r}/2,\tau_{2})\right)\right\} .\nonumber 
\end{eqnarray}
From this equation we see that $F(\vec{r},\vec{p}^{\,\prime}-\vec{p}=0)=R^{4}E^{{\rm dir}}(r)$.
For small $r\rho^{-1}$ we may apply an electric dipole approximation
during the calculations of $F(\vec{r},\vec{p}^{\,\prime}-\vec{p})$.
As we already mentioned, the dipole approximation may be well-justified
 in the botomonium case, while for the charmonium we expect sizable
corrections from other terms of the expansion.

\subsection{Instantons and heavy quarkonia's sizes}

\label{QQsizes}

In order to understand the applicability of ILM in the heavy quark
sector, we should compare the typical sizes of quarkonia and ILM model
parameters. For example, the sizes of heavy quarkonia are relatively
small~\cite{Digal:2005ht,Eichten:1979ms} (see Table~\ref{Quarkoniumsizes}).
One can see, that this is more pronounced for the low-lying states
of sizes $r_{J/\psi}$ and $r_{\Upsilon}$.

\begin{table}[h]
\begin{centering}
\caption{Masses and sizes of quarkonium states in the non-relativistic potential
model \cite{Digal:2005ht}.}
\par\end{centering}
\scalebox{1}{ %
\begin{tabular}{|c|c|c|c|c|c|c|c|c|}
\hline 
Characteristics  & \multicolumn{3}{c|}{Charmonia} & \multicolumn{5}{c|}{Bottomonia}\tabularnewline
\hline 
of states  & $J/\psi$  & $\chi_{c}$  & $\psi'$  & $\Upsilon$  & $\chi_{b}$  & $\Upsilon'$  & $\chi_{b}'$  & $\Upsilon^{''}$ \tabularnewline
\hline 
mass [GeV]  & 3.07  & 3.53  & 3.68  & 9.46  & 9.99  & 10.02  & 10.26  & 10.36 \tabularnewline
\hline 
size $r$ [fm]  & 0.25  & 0.36  & 0.45  & 0.14  & 0.22  & 0.28  & 0.34  & 0.39 \tabularnewline
\hline 
\end{tabular}} \label{Quarkoniumsizes} 
\end{table}

Estimates of nucleon's quark core sizes also give the similar results
$r_{N}\sim0.3-0.5$\,fm~\cite{He:1986yq,Weise,Tegen}. While the
quark cores of hadrons are relatively small, one may conclude that
they are insensitive to the confinement mechanism which is pronounced
at distances $\sim1$\,fm. Consequently, the ILM may be safely applied
for the description of hadron properties at the heavy quark sector
too. During this applications one can apply a systematic approach
to take into account the nonperturbative effects in the hadron properties
in terms of the packing parameter $\kappa$. However, the perturbative
effects must be carefully taken into account during the analysis of
heavy quarkonia spectra and their wave functions.

\subsection{Standard approach and Phenomenology of the $(Q\bar{Q})_{n'}\rightarrow(Q\bar{Q})_{n}\,\,\pi\pi$
process}

\label{QQpipi-standard} According to Ref.\,\cite{Mannel1997} the
phenomenological definition of the coupling for $(Q\bar{Q})_{n'}\rightarrow(Q\bar{Q})_{n}\,\pi\pi$
process can be written in the form of the effective lagrangian ${\cal L}$.
In the chiral $(m_{q}\rightarrow0)$ and heavy quark mass $(m_{Q}\rightarrow\infty)$
limits it has the form 
\begin{eqnarray}
{\cal L}=gA_{\mu}^{(v)}B^{(v)\mu*}{\rm tr}[(\partial_{\nu}U)(\partial^{\nu}U)^{\dagger}]+h.c.
\end{eqnarray}
where $A_{\mu}^{(v)}$ and $B^{(v)\mu}$ are factors corresponding
to $(Q\bar{Q})_{n'}\sim(2S)$ and $(Q\bar{Q})_{n}\sim(1S)$ states.
The experimental values of the couplings are given in Table\,\ref{Table4}.
\begin{table}[h]
{\footnotesize{}\caption{The values of coupling $g$ from the decay processes.}
}{\footnotesize\par}
\begin{centering}
{\footnotesize{}}%
\begin{tabular}{@{}ccc}
\hline 
 & {\footnotesize{}$\psi(2S)\rightarrow J/\psi\pi^{+}\pi^{-}$ } & {\footnotesize{}$\Upsilon(2S)\rightarrow\Upsilon(1S)\pi^{+}\pi^{-}$}\tabularnewline
\hline 
{\footnotesize{}g } & {\footnotesize{}$0.30\pm0.02$ } & {\footnotesize{}$0.25\pm0.02$ }\tabularnewline
\hline 
\end{tabular}{\footnotesize\par}
\par\end{centering}
{\footnotesize{}\label{Table4} }{\footnotesize\par}
\end{table}

Our estimate of $g_{J/\psi}$ in the framework of ILM is 
\begin{eqnarray}
g_{J/\psi}=\frac{F_{\pi Q}^{2}}{F_{\pi}^{2}}1.345\frac{r_{J/\psi}^{2}}{\rho^{2}}\left(1-0.372\frac{r_{J/\psi}^{2}}{\rho^{2}}\right)\label{g}
\end{eqnarray}
where the quantity outside of the bracket corresponds to the dipole
approximation, while the bracket takes into account the higher order
corrections in the expansion over $(r/\rho)^{2}$. As expected, it
is seen that $g_{\Upsilon}<g_{J/\psi}$.

\begin{figure}[h]
\begin{centering}
\includegraphics[scale=0.5]{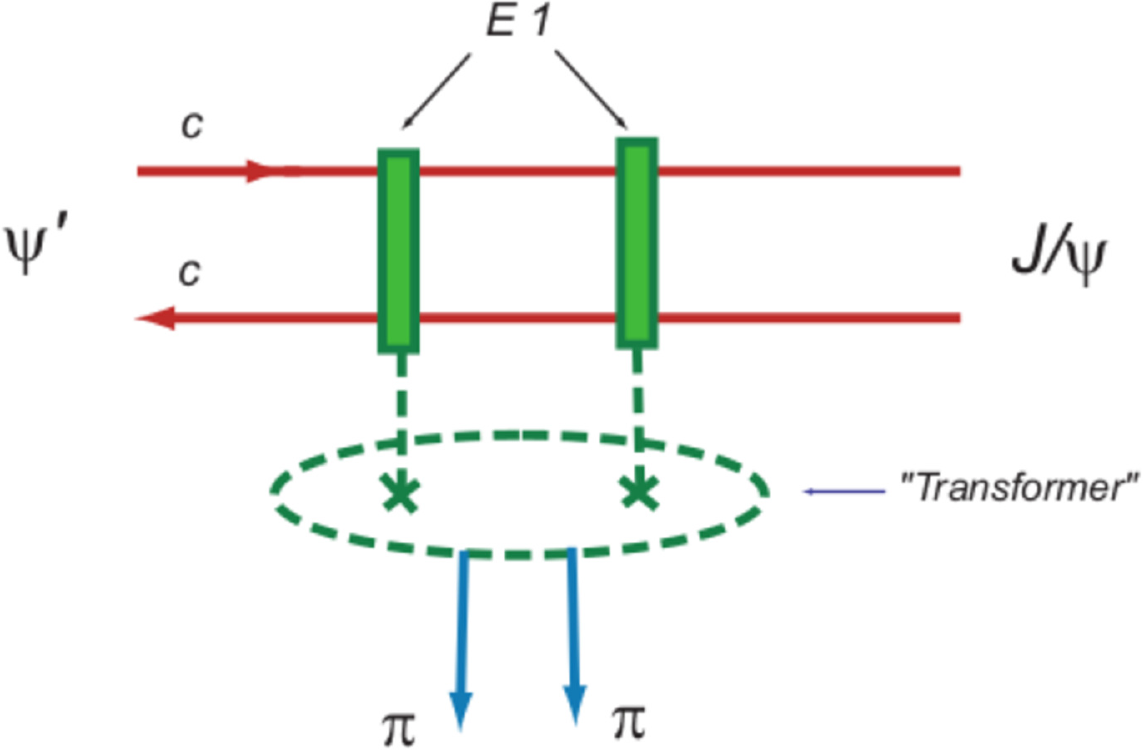}
\par\end{centering}
\caption{Two pion production from the charmonia transitions in dipole approximation.}
\label{Fig8} 
\end{figure}

The standard approach to `the quarkonium -- light hadron transitions'
assumes an applicability of the multipole expansion, which means that
the quarkonium sizes $r_{c,b}$ are much less than the typical size
of the nonperturbative vacuum gluon fluctuation $\lambda_{g}$ (see
e.g.~\cite{Voloshin:2007dx}). According to this assumption, the
dipole approximation can be represented as shown in Fig.\,\ref{Fig8}.
However, in ILM $\lambda_{g}\approx\rho$. Using the instanton parameter
$\rho=0.33$~fm and the charmonium size $r_{J/\psi}=0.25$~fm (see
Table\,\ref{Quarkoniumsizes}) from Eq.\,(\ref{g}) we can get the
value $g_{J/\psi}=0.28(1-0.2)\simeq0.22$. From here we can conclude
that the correction to the dipole approximation in charmonium case
is quite sizable, $\sim20\%$.

\section{Heavy quark correlators with perturbative corrections in ILM}

\label{QQCor}

Our previous discussion demonstrated the importance of the nonpertubative
contributions for the spectra and wave functions of heavy quarkonia,
which might be naturally incorporrated if we will use $Q\bar{Q}$
potential found within ILM approach. We will follow our previous work~\cite{Musakhanov:2020hvk}.
In calculations, one should also take into account  that the instanton
field has a specific $A\sim1/g$ dependence on the strong coupling
$g$. The normalized partition function $Z[j]$  in ILM can be given
by an approximate expression 
\begin{eqnarray}
Z[j] & =\int D\xi Dae^{-[S_{eff}[a,A(\xi)]+(ja)]}\approx\int D\xi e^{-\frac{1}{2}(j_{\mu}S_{\mu\nu}(\xi)j_{\nu})},\label{Z1}
\end{eqnarray}
which accounts for the perturbative gluons $a_{\mu}$ and their corresponding
sources $j_{\mu}$. In derrivation of the partition function in Eq.\,(\ref{Z1}),
we neglected the self-interaction terms of order  ${\cal O}(a^{3},a^{4})$
, and  used the shorthand notations 
\begin{eqnarray*}
(ja)=\int d^{4}xj_{\mu}^{a}(x)a_{\mu}^{a}(x),\,\,\,
j_{\mu}^{a}(x)S_{\mu\nu}^{ab}(x,y,\xi)j_{\nu}^{b}(y).
\end{eqnarray*}
where $S_{\mu\nu}^{ab}(x,y,\xi)$ is a gluon propagator in the presence
of the instanton background $A(\xi)$.  Since an infinitely heavy
quark interacts only through the fourth components of instantons $A_{4}$
and perturbative gluon $a_{4}$ fields, respectively, we need only
$S_{44}(\xi)$ components of a gluon propagator. 

Now we will focus on analysis of the heavy quark propagator in ILM
framework. From Eq.~(\ref{Z1}) it is seen that the averaged heavy
quark propagator ${w}$ with the account of perturbative gluon field
fluctuations $a$ is given by the expression 
\begin{eqnarray}
{w} & =\int D\xi Da\exp[-S_{eff}(a,\xi)+(ja)]\left(\theta^{-1}-ga-g\sum_{i}A_{i}\right)^{-1}\\
 & =\int D\xi\left[\int\left(\theta^{-1}-g\frac{\delta}{\delta j}-g\sum_{i}A_{i}\right)^{-1}%
\exp\left\{ \frac{1}{2}(jS(\xi)j)\right\} \right]_{j=0}.\label{w1}
\end{eqnarray}
It is straightforward to prove that 
\begin{eqnarray}
\left[\frac{1}{\theta^{-1}-g\frac{\delta}{\delta j}-gA(\xi)}\exp\left(\frac{1}{2}jS(\xi)j\right)\right]_{_{j=0}}=\left[\exp\left(\frac{1}{2}\frac{\delta}{\delta a_{a}}S_{ab}(\xi)\frac{\delta}{\delta a_{b}}\right)\frac{1}{\theta^{-1}-ga-gA(\xi)}\right]_{_{a=0}}\qquad\label{q1}
\end{eqnarray}
Furthermore, this equation can be extended to any correlator. Consequently,
the path integral of heavy quark functional $F[A(\xi),a]$ in the
approximations discussed above can be given by the  equation 
\begin{eqnarray}
\int D\xi Da\exp\left\{ -S_{eff}[A(\xi),a]\right\} F[A(\xi),a]=\int D\xi\left[\exp\left(\frac{1}{2}\frac{\delta}{\delta a_{a}}S_{ab}(\xi)\frac{\delta}{\delta a_{b}}\right)F[\xi,a]\right]_{a=0}.\label{F}
\end{eqnarray}
Another equation similar to this,  in the absence of instanton background
$A(\xi)=0$ and for the gluon propagator taken in Coulomb gauge, was
suggested before in Ref.~\cite{brown1979}.

As we mentioned at the subsection~\ref{ILM}, the systematic accounting
of the nonperturbative effects in the ILM can be performed in terms
of the dimensionless parameter $\kappa$ by using the Pobylitsa equations~\cite{Pobylitsa:1989uq}.
The situation here is quite comfortable for the performing systematic
analysis of instanton effects, since  the $\kappa$ value is very
small at the values of instanton parameters discussed above, $\kappa\sim0.01$.

In order to take into account the perturbative gluons effects, we
should perform an expansion in terms of parameter $\alpha_{s}$. As
was discussed in the subsection~\ref{ILM}, in ILM on its scale $\bar{\rho}\approx0.33$~fm
we have $\alpha_{s}(\rho)\approx0.39$. The pure perturbative effects
at the leading order appear as linear corrections in $\alpha_{s}$.
A systematic analysis, including  both the perturbative and nonperturbative
effects, requires a double  series expansion in terms of $\alpha_{s}$
and $\kappa$. In order to perform such  analysis, we may assume that
$\alpha_{s}\sim\kappa^{1/2}$, which is quite reasonable according
to the phenomenological studies. Consequently, during the calculations
one should keep all necessary terms at the order of ${\cal O}(\kappa)$
and ${\cal O}(\alpha_{s}\kappa^{1/2})$.

\subsection{Gluons in ILM}

\label{gluons}

In the above mentioned approximation, the gluon propagator in the
instanton medium can be represented by re-scattering series as 
\[
S(\xi)=S^{0}+\sum_{i}\Delta S^{i}(\xi_{i})+...,\quad\Delta S^{i}(\xi_{i})\equiv S^{i}(\xi_{i})-S^{0},
\]
where $S^{0}$ is free gluon propagator, and $S^{i}(\xi_{i})$ is
propagator of gluon in the instanton background. The averaged value
of gluon propagator $\overline{S}$ in ILM can be found by extending
the Pobylitsa's equation to the gluon case~\cite{Musakhanov:2017erp}
\begin{eqnarray}
\overline{S}(k)=\frac{1}{k^{2}+M_{g}^{2}(k)}.\label{glprop}
\end{eqnarray}
Consequently, the perturbative gluons are also acquire the momentum
dependent mass, which  is defined by the  expressions 
\begin{eqnarray}
M_{g}(k) & =M_{g}(0)F(k),\quad M_{g}(0)=\frac{2\pi}{\rho}\left(\frac{6\kappa}{N_{c}^{2}-1}\right)^{1/2},\quad F(k)=k\rho K_{1}(k\rho).\label{Mg0}
\end{eqnarray}
Here $K_{1}$ is a modified Bessel function of the second type. Using
the typical values of instanton parameters $\rho=1/3\,{\rm fm},\,\,R=1\,{\rm fm}$,
we can estimate the dynamical gluon mass at zero momentum. Its value
is given by $M_{g}(0)\simeq358\,{\rm MeV}$ and is close to the value
of dynamical light quark mass. We can also note, that the dynamical
gluon and light quark masses appear at the same order  ${\cal O}\big(\kappa^{1/2}\rho^{-1}\big)$.
The gauge invariance of the dynamical gluon mass $M_{g}$ was proven
in Ref.~\cite{Musakhanov:2017erp}.

One may wonder if the instantons also generate the nonperturbative
gluon-gluon interactions and, in such a way, contribute to the glueballs'
properties. The corresponding investigations in instanton liquid model~\cite{Schafer:1994fd,Tichy:2007fk}
devoted to the $J^{PC}=0^{++},0^{-+},2^{++}$ glueballs, showed that
the instanton-induced forces between gluons  lead to the strong attraction
in the $0^{++}$ channel, to the strong repulsion in the $0^{-+}$
channel and to the absence of short-distance effects in the $2^{++}$
channel. Consequently, applications of ILM in studies of glueballs
predicted hierarchy of the masses $m_{0^{++}}<m_{2^{++}}<m_{0^{-+}}$
and their corresponding sizes $r_{0^{++}}<r_{2^{++}}<r_{0^{-+}}$.
These predictions were confirmed by the lattice calculations~\cite{deForcrand:1991kc,Weingarten:1994vc,Chen:1994uw,Morningstar:1999rf,Athenodorou:2020ani,Meyer:2004jc,Meyer:2004gx}.
For typical values of ILM parameters $\rho=1/3$~fm and $R=1$~fm
it has been found~\cite{Schafer:1994fd} that the mass of $0^{++}$
glueball $m_{0^{++}}=1.4\pm0.2$~GeV and its size $r_{0^{++}}\approx0.2$~fm,
in a nice correspondence with the lattice calculations~\cite{deForcrand:1991kc,Weingarten:1994vc,Chen:1994uw}.
Further studies of the $0^{++}$ glueball in ILM~\cite{Tichy:2007fk}
gave $m_{0^{++}}=1.29\,-\,1.42$~GeV, which was also in a good agreement
with the lattice results~\cite{Meyer:2004jc,Meyer:2004gx}.

Main conclusion of the studies which we discussed above was that the
origin of $0^{++}$ glueball is mostly provided by the short-sized
nonperturbative fluctuations (instantons), rather than the confining
forces. In a quick summary, we may conclude that ILM provides the
promising framework for describing the lowest state glueball's properties.

\subsection{Heavy quark propagator with perturbative corrections in ILM}

\label{Qprop}

Let us now discuss a single heavy quark properties in ILM by estimating
the corresponding effects from perturbative and nonperturbative regions,
as  was done in Ref.\,\cite{Musakhanov:2020hvk}. An averaged infinitely
heavy quark propagator in ILM according to Eqs.(\ref{w1})-(\ref{q1})
is given by 
\begin{eqnarray}
{w}=\!\left.\!\int\!\!D\xi\exp\left[\frac{1}{2}\!\left(\frac{\delta}{\delta a}S(\xi)\frac{\delta}{\delta a}\right)\!\right]\!\frac{1}{\theta^{-1}-ga-gA(\xi)}\right|_{a=0}\!\!\!\label{wQpert}
\end{eqnarray}
where we have used the shorthand notation
\begin{equation}
\left(\frac{\delta}{\delta a}S(\xi)\frac{\delta}{\delta a}\right)=\int d^{4}yd^{4}z\frac{\delta}{\delta a_{a}(y)}S_{ab}(\xi,y,z)\frac{\delta}{\delta a_{b}(z)}.
\end{equation}
From our work~\cite{Musakhanov:2020hvk} we can see that in the ILM
the heavy quark propagator with perturbative corrections can be written
as 
\begin{equation}
{w}=\int D\xi\left[\theta^{-1}-\sum_{i}\left(gA_{i}(\xi_{i})-g^{2}\left(\Delta S^{i}(\xi_{i})\theta\right)\right)\right]^{-1},\label{w11}
\end{equation}
where the last term in the denominator means the heavy quark mass
operator, which is formally a correction of the order ${\cal O}(\alpha_{s}\kappa^{1/2})$.
Heavy quark propagator Eq.~(\ref{w11}) and its special $g\rightarrow0$
limit expression have similar structures, according to their dependencies
on the instanton collective coordinates. We can now extend Pobylitsa's
equation from Ref.~\cite{Diakonov:1989un}, which in the approximation
of small ${\cal O}(\kappa,\,\alpha_{s}\kappa^{1/2})$ has a form 
\begin{eqnarray}
{w}^{-1} & = &\nonumber\theta^{-1}-\sum_{i}\int\xi_{i}\theta^{-1}\left(\frac{1}{\theta^{-1}-gA_{i}(\xi_{i})}-\theta\right)\theta^{-1}-g^{2}\left((\bar{S}-S^{0})\theta\right).\label{wQ-1pert}
\end{eqnarray}
In the last term the averaged gluon propagator $\bar{S}$ is given
by Eq.~(\ref{glprop}). In such a way, the second term in the right
side of Eq.~(\ref{w-1}) leads to the ILM heavy quark mass shift
$\Delta M_{Q}^{{\rm dir}}$ with the corresponding order ${\cal O}(\kappa),$
while the third term is ILM-modified perturbative gluon contribution
to the heavy quark mass $\Delta M_{Q}^{{\rm pert}}$, which is formally
of order  ${\cal O}(\alpha_{s}\kappa^{1/2})$, respectively. We note
that the direct mass contribution to the quark mass in instanton background
$\Delta M_{Q}^{{\rm pert}}$ was calculated first in Ref.~\cite{Diakonov:1989un}.
Numerically, we can estimate that in ILM 
\begin{eqnarray}
-\Delta M_{Q}^{{\rm pert}}\sim\frac{2}{N_{c}}\,\alpha_{s}M_{g}(0)\sim\Delta M_{Q}^{{\rm dir}}\simeq70\,{\rm MeV}.\label{DeltaM-pert}
\end{eqnarray}
We see that ILM perturbative $\Delta M_{Q}^{{\rm pert}}$ and ILM
direct $\Delta M_{Q}^{{\rm dir}}$ corrections to the heavy quark
mass almost completely cancel each other. This estimate is in agreement
with the above made assumptions ${\cal O}(\alpha_{s}\kappa^{1/2})\sim{\cal O}(\kappa)$.

\subsection{Heavy quarks  singlet potential with perturbative corrections in
ILM}

\label{QQpotential}

Heavy quarks $Q\bar{Q}$ correlator in ILM can be calculated from
the operator 
\begin{eqnarray}
{W}=\int D\xi\exp\left[\frac{1}{2}\sum_{i,j=1}^{2}\left(\frac{\delta}{\delta a_{a}^{(i)}}S_{ab}^{(ij)}(\xi)\frac{\delta}{\delta a_{b}^{(j)}}\right)\right]\left.\frac{1}{D^{(1)}-ga^{(1)}}\frac{1}{\bar{D}^{(2)}-g\bar{a}^{(2)}}\right|_{a=0},\label{W1}
\end{eqnarray}
where the operator $D^{(1)}$ is defined as $D^{(1)}=\theta^{-1}-gA^{(1)}(\xi)$
and, $a^{(1)}$ and $A^{(1)}$ are the corresponding fields projections
to the line $L_{1}$, respectively. Similarly, one has $\bar{D}^{(2)}=\theta^{-1}-g\bar{A}^{(2)}(\xi)$
where $\bar{a}^{(2)}$ and $\bar{A}^{(2)}$ are the corresponding
fields projections to the line $L_{2}$, respectively (see Fig.~\ref{W-loop}).
Consequently, extended Pobylitsa's equation in the approximation ${\cal O}(\kappa,\alpha_{s}\kappa^{1/2})$
has the solution~\cite{Musakhanov:2020hvk}: 
\begin{eqnarray}
 &  & {W}^{-1}={w}^{(1)-1}{\bar{w}}^{(2)-1}-\sum_{i}\int d\xi_{i}\theta^{(1)-1}\\
 &  & \times\left(\frac{1}{D_{i}^{(1)}}-\theta^{(1)}\right)\theta^{(1)-1}\theta^{(2)-1}\left(\frac{1}{\bar{D}_{i}^{(2)}}-\theta^{(2)}\right)\theta^{(2)-1}-g^{2}\frac{\lambda_{a}}{2}\frac{\bar{\lambda}_{b}}{2}\int D\xi\,S_{ab}^{(12)},\label{W-1}
\end{eqnarray}
where ${w}^{(1)-1}$ is given by Eq.(\ref{w-1}). Note, that   the
equation for ${\bar{w}}^{(2)-1}$ also can be written in a form similar
to Eq.(\ref{w-1}).

\begin{figure}[hbt]
\begin{centering}
\includegraphics[scale=0.75]{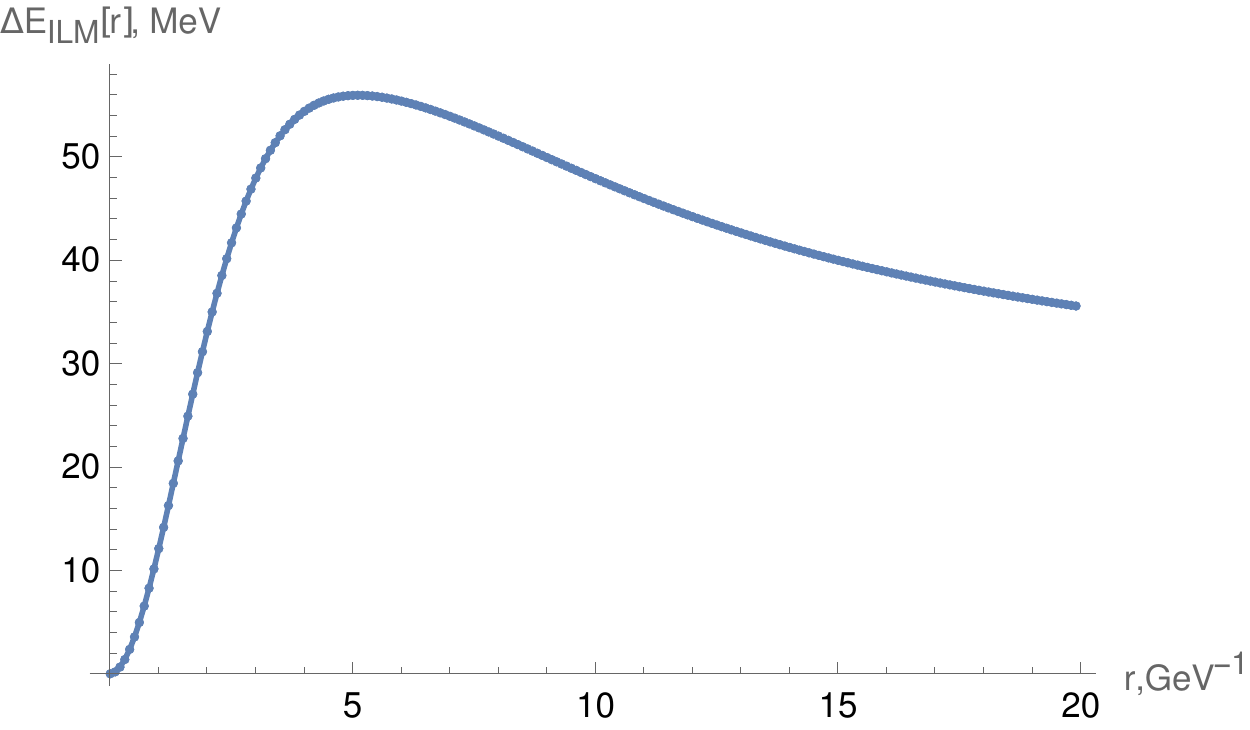}
\par\end{centering}
\caption{ILM total contribution $\Delta E_{{\rm ILM}}(r)$ to the energy of
singlet $Q\bar{Q}$ system as a function of their relative distance.
Basic parameters are taken as: $R=1$~fm, $\rho=0.33$~fm, $\alpha_{s}(\rho)=0.39$.}
\label{DeltaE} 
\end{figure}

It is clear that ILM contributions to the heavy quark masses $\Delta M_{Q}^{{\rm dir}}$
and $\Delta M_{Q}^{{\rm pert}}$ are coming from the first term of
the Eq.~\re{W-1}, while $V_{{\rm dir}}(r)$ - from second, and
$V_{{\rm pert}}(r)$ - from third one. First we note that 
\begin{eqnarray}
 &  & V_{{\rm pert}}(r)=V_{C}(r)+\Delta V_{{\rm pert}}(r),\,\,V_{C}(r)=-\frac{16\pi\alpha_{s}}{3}\int\frac{d^{3}q}{(2\pi)^{3}}e^{i\vec{q}\vec{r}}\frac{1}{q^{2}},\,\,\,\nonumber \\
 &  & \Delta V_{{\rm pert}}(r)=\frac{16\pi\alpha_{s}}{3}\int\frac{d^{3}q}{(2\pi)^{3}}e^{i\vec{q}\vec{r}}\frac{M^{2}(q)}{q^{2}(q^{2}+M_{g}^{2}(q))},\,\,\,q=|\vec{q}|,\label{DeltaVpert}
\end{eqnarray}
where $M_{g}(q)$ is given by Eq.~(\ref{Mg0}), $V_{C}(r)$ is the
perturbative QCD one gluon exchange (OGE) potential and $\Delta V_{{\rm pert}}(r)$
is ILM contribution $\sim O(\kappa^{1/2}\alpha_{s})$, which is positively
defined. We can find now from the Eq.~\re{W-1} the ILM contribution
to the total energy of $Q\bar{Q}$ system in singlet color state as
\begin{eqnarray}
\Delta E_{{\rm ILM}}(r)=\Delta E_{{\rm ILM}}^{{\rm dir}}(r)+\Delta E_{{\rm ILM}}^{{\rm pert}}(r),\,\,\,\Delta E_{{\rm ILM}}^{{\rm dir}}(r)=2\Delta M_{Q}^{{\rm dir}}+V_{{\rm dir}}(r),\,\,\,\Delta E_{{\rm ILM}}^{{\rm pert}}(r)=2\Delta M_{Q}^{{\rm pert}}+\Delta V_{{\rm pert}}(r)
\end{eqnarray}
The direct instanton contribution $\Delta E^{{\rm dir}}(r)$ was calculated
early in Ref.~\cite{Diakonov:1989un}. It is clear that $\Delta E^{{\rm dir}}(r)$
is positively defined, since $\Delta E^{{\rm dir}}(0)=0$ and $\Delta E^{{\rm dir}}(\infty)=2\Delta M_{Q}^{{\rm dir}}>0$.
On the other hand, comparing Eqs.~(\ref{DeltaM-pert}) and (\ref{DeltaVpert}),
we find that $\Delta E_{{\rm ILM}}^{{\rm pert}}(r)$ is negatively
defined, since $\Delta E_{{\rm ILM}}^{{\rm pert}}(0)=0$ and $\Delta E_{{\rm ILM}}^{{\rm pert}}(\infty)=2\Delta M_{Q}^{{\rm pert}}<0$.
So, the two contributions almost cancel each other, and the total
ILM contribution $\Delta E_{{\rm ILM}}(r)$ becomes small, as could
be seen from the Fig.~\ref{DeltaE}. For this reason we expect small
influence of the instantons on the heavy quarkonia spectra and wave
functions.

The relative sizes of all possible instanton effects in comparison
with the results corresponding to the Cornell's model results~\cite{Mateu:2018zym}
found to be few percents in the ILM~\cite{Turimov:2016adx}.

\section{Summary}

\label{summary} QCD vacuum instantons generate the nonperturbative
interactions among gluons, light and heavy quarks. The phase state
of the instanton ensemble in the framework of the Instanton Liquid
Model (ILM) is controlled by the packing parameter $\kappa=\rho^{4}/R^{4}$,
which quantifies the fraction of the space occupied by instantons.
Detailed considerations of the instanton size $\rho$ and inter-instanton
distance $R$ showed that in average $\bar{\rho}\approx1/3$~fm,
and $R\approx1$~fm, so the packing parameter is comfortably small
$\kappa\approx0.01$. This finding can be used as expansion parameter
within Pobylitsa equations for  averaging over the instanton collective
coordinates in various correlators. Most essential conclusions can
be made by considering the strength of particle-instanton interaction.
We found that the dominance of zero-modes in light quark determinant
provides strong light quark-instanton interaction, which leads to
dynamical light quark mass $M_{q}\approx360$~MeV$\sim\kappa^{1/2}\rho^{-1}$.
In contrast, for heavy quarks their interaction with gluons is controlled
mainly by their mass, and for this reason they rather weakly interact
with instantons, so the instanton contribution to their mass is small
and given by $\Delta M_{Q}^{{\rm dir}}\approx70$~MeV$\sim\kappa\rho^{-1}.$
 It is clear that instantons generate nonperturbative interactions
between particles, which cross simultaneously the same instanton.
This mechanism  perfectly describes spontaneous breaking of chiral
symmetry (SBChS), which is most essential phenomena for light quarks
physics. On the other hand, this nonperturbative mechanism also contributes
to the interaction between light and heavy quarks and might be used
for studies of nonperturbative effects in light-heavy quarks systems.
The manuscript provides the framework for the calculations of various
problems for such a systems. As an example, we considered in detail
the process of pion pair emission in decays of excited heavy quarkonia.
This analysis requires  calculation of heavy quarkonia spectra and
their wave functions. While usually this analysis is done using perturbative
QCD, which  describes them fairly well, in our study we calculated
ILM corrections to these quantities. In the calculations we took into
account the perturbative gluon properties in the instanton medium.
We found that perturbative gluon-instanton interaction is strong,
as happens for light quarks, and the instanton-generated dynamical
gluon mass $M_{g}\approx360$~MeV$\sim\kappa^{1/2}\rho^{-1}$ is
comparable to that of the light quarks. Finally, we found that $O(\kappa)$
contributions to the singlet heavy $Q\bar{Q}$ system energy almost
completely cancel similar $O(\alpha_{s}\kappa^{1/2})$ contributions,
and thus nonperturbative interactions have small influence on  the
heavy quarkonia spectrum and wave functions. In our evaluation we
neglect possible contributions of the higher Fock state $Q\bar{q}q\bar{Q}$,
which might be relevant at large scale distances $\sim 1/M_q\sim0.5$~fm. We
are planning to calculate this and similar problems related to the
heavy-light quarks systems within ILM.

 \section*{Acknowledgments}
I am grateful to Marat Siddikov for his valuable comments and suggestions.

\end{document}